\RequirePackage[l2tabu, orthodox]{nag} 

\RequirePackage{fix-cm} 

\documentclass{birkjour}

\usepackage[hyphens]{url}
\usepackage{hyperref}
\hypersetup{breaklinks=true}

\usepackage{complexity}

\usepackage{amssymb} 
\usepackage{amsfonts} 
\usepackage{amsthm} 
\usepackage{amsmath} 

\usepackage[ruled,vlined,linesnumbered]{algorithm2e}
\SetKwProg{Fn}{Function}{}{}
\newcommand{\funct}[1]{$\mathtt{#1}$}
\usepackage{enumitem} 
 
 \usepackage{mathpazo}

 \newtheorem{thm}{Theorem}[section]
 
 \newtheorem{lem}[thm]{Lemma}
 \newtheorem{prop}[thm]{Proposition}
 
 \theoremstyle{definition}
 \newtheorem{defn}[thm]{Definition}
 \theoremstyle{remark}
\newtheorem*{rem}{\bf Remark}

\usepackage{mathrsfs}

\usepackage{etoolbox}

\usepackage{microtype}

\usepackage{booktabs}

\usepackage[]{nicefrac}

\usepackage{nth}

\usepackage{paralist}

\usepackage{markdown}

\usepackage{float}

\usepackage{subcaption}

\usepackage{xspace}

\usepackage{listings}
\newbool{use_todo}
\setbool{use_todo}{true}
\ifbool{use_todo}{
	\usepackage[colorinlistoftodos, shadow, textsize=small, textwidth=35mm]{todonotes}
	\setlength{\marginparwidth}{4.0cm}
	}{
	\usepackage[disable]{todonotes}
}

\definecolor{forestgreen}{rgb}{0.13, 0.55, 0.13}

\definecolor{deepblue}{rgb}{0,0,0.5}
\definecolor{deepred}{rgb}{0.6,0,0}
\definecolor{deepgreen}{rgb}{0,0.5,0}
\definecolor{verydarkGray}{RGB}{30,30,30}
\definecolor{darkGray}{RGB}{80,80,80}
\definecolor{midGray}{RGB}{160,160,160}  
\definecolor{midBlue}{RGB}{120,120,160}  
\definecolor{verylightgray}{RGB}{210,210,210}
\definecolor{verylightred}{rgb}{0.8,0.1,0.1}
\definecolor{gray}{rgb}{0.4,0.4,0.4}
\definecolor{darkblue}{rgb}{0.0,0.0,0.6}
\definecolor{cyan}{rgb}{0.0,0.6,0.6}
\definecolor{forestGreen}{RGB}{0, 153, 76}

\begin{document}

\title[Computational aspect of GA products]{Computational Aspects of Geometric Algebra Products of Two Homogeneous Multivectors}

\author{Stephane Breuils}
\address{National Institute of Informatics, Tokyo 101-8430, Japan}
\email{breuils@nii.ac.jp}

\author{Vincent Nozick}
\address{Laboratoire d'Informatique Gaspard-Monge, Equipe A3SI,\\ UMR 8049, Universit\'e Paris-Est Marne-la-Vall\'ee, France}
\email{vincent.nozick@u-pem.fr}

\author{Akihiro Sugimoto}
\address{National Institute of Informatics, Tokyo 101-8430, Japan}
\email{sugimoto@nii.ac.jp}

\subjclass{Primary 99Z99; Secondary 00A00}

\keywords{Geometric Algebra, Clifford Algebra, Computational complexity, Arithmetic operations}

\date{October, 2019}

\begin{abstract}
    Studies on time and memory costs of products in geometric algebra have been limited to cases where multivectors with multiple grades have only non-zero elements. This allows to design efficient algorithms for a generic purpose; however, it does not reflect the practical usage of geometric algebra. Indeed, in applications related to geometry, multivectors are likely to be full homogeneous, having their non-zero elements over a single grade. 
    In this paper, we provide a complete computational study on geometric algebra products of two full homogeneous multivectors, that is, the outer, inner, and geometric products of two full homogeneous multivectors. We show tight bounds on the number of the arithmetic operations required for these products. We also show that algorithms exist that achieve this number of arithmetic operations.  
\end{abstract}

\maketitle



\section{Introduction}
\label{section:introduction}


Geometric algebra presents intuitive solutions for problems related to geometry. Its theory is more and more investigated in various research fields like physics, mathematics or computational geometry, see~\cite{Kanatani2015,Lasenby2019,Keni2019} for some examples. In contrast, in the computer science field, the study of computational aspects of the geometric algebra operators is still limited. It has started thanks to the pioneering work of \cite{Fon07}, which gave some results about complexity of geometric algebra products in the worst case. The worst case here means all the elements of a multivector with multiple grades are non-zero. Their study is based on the most used approach to deal with products in geometric algebras. Namely, the approach deals with fast binary indices and per-bit operators like the XOR operator~\cite{fontijne2010gaviewer}. 
The most used algorithms deal with fast binary indices and per-bit operators like XOR operator, which are used to generate products, see~\cite{colapinto2011versor,LeoBook,Fon07} or GATL in~\cite{gatlImplementation} for example. Hereafter, we refer to such approaches as the XOR method.



The XOR algorithm of the products consists of representing the basis blades with binary indices and computing the products with logical operators between these indices.   
For each product, the XOR algorithm runs first in looping over two multivectors. The computational cost of this operation is $\mathcal{O} ( 4^d ) $, where $d$ is the dimension of the vector space. In this kind of algorithm, the computation of the sign is evaluated by computing a ``convolution'' between two binary indices. The convolution consists of right-shifting each bit of one index until it is zero. At each iteration,  the number of ones in common between the shifted index and the other index is counted. The sign is obtained by raising $-1$ to the power of the number of ones. The computational cost of this operation is linear to the dimension. Indeed, for any grades, the number of right-shifting is $d$, leading to the complexity of $\mathcal{O}(d \times 4^d )$, see~\cite{breuils_garamon_2019}. 

\subsection{Full homogeneous multivectors}
Multivectors are, in its practical usage, likely to be homogeneous, i.e., have their non-zero elements concentrated in a single grade, see, for example, the representation of any geometric objects in CGA~\cite{LeoBook}. 
This paper deals only with homogeneous multivectors. This assumption does not limit the scope of this paper. Indeed, in case the multivector is not homogeneous, then it is still defined as the sum of homogeneous multivectors. Furthermore, the operators of geometric algebra are distributive with respect to the addition. Then,any products between non-homogeneous multivectors can be reduced to some products of homogeneous multivectors.
We also assume that all the elements of the homogeneous multivectors are non-zero. In most applications dealing with geometry such as~\cite{Kanatani2015,DietmarBook,benger2014differential}, the multivectors contain only non-zero elements. 
We call a multivector with only a single grade and having non-zero coefficients only, a full homogeneous multivector.
We remark that although some non full homogeneous multivectors exist (see~\cite{yuan2014clifford}), dealing with them is out of the scope of this paper.

Over the full homogeneous multivectors, we focus on only three operators, namely, the outer product denoted by~`$\wedge$', the inner product denoted by~`$\cdot$' and the geometric product denoted by~`$*$'. There exist more operators such as the dual, inverse, see~\cite{LeoBook} for a more exhaustive list. However, all these operators can be obtained from the three aforementioned operators, see~\cite{hestenes1996grassmann}.

\subsection{Notation}
\label{sec:notation}
Following the state-of-the-art usages in~\cite{LeoBook} and~\cite{perwassGeometric}, lower-case bold letters refer to vectors (vector $\mathbf{a}$) and lower-case non-bold to multivector coordinates (coefficient $a_i$). Upper-case bold letters denote blades (blade~$\mathbf{A}$) whose grade is higher than $1$. Multivectors and $k$-vectors are denoted with upper-case non-bold letters (multivector~$A$). Lower-case and Frakture letters denote multivector expressed over a tree structure. For example, $\mathfrak{a}$~represents a multivector over a tree structure, this notion is detailed in~Section~\ref{sec:recursiveApproach}.
The part of grade~$k$ of a multivector $A$ is denoted by $\langle A \rangle_{k}$. The total number of basis blades is $2^d$, where~$d$ is the number of basis blades $\mathbf{e}_{i}$ of grade~$1$. 
Throughout this paper, a basis blade of grade $k$ will be denoted using set theory notation. To achieve this, we will assume an orthogonal basis called $\mathcal{B}=\{ \mathbf{e}_1,\mathbf{e}_2,\dots ,\mathbf{e}_d  \}$, with $d$ (the vector space dimension). Similarly to the notation in~\cite{perwassGeometric}, a basis blade of grade $g$ is denoted by
\begin{equation}
    \mathbf{e}_{\{ \mu \}} = \mathbf{e}_{ \mu_1 } \wedge \mathbf{e}_{ \mu_2 } \wedge \cdots \wedge \mathbf{e}_{ \mu_g } \text{,~~~ where }\mu = \{ \mu_1, \mu_2, \dots, \mu_g\}
\end{equation}
with a greek letter as the subscript. We note that $\mu \subseteq P(\mathcal{B})$ where $P$ is the power set (the set of all the subsets of a set). 
For example, the blade $\mathbf{e}_{234}$ can be referred to as $\mathbf{e}_{\mu}$ with $\mu = \text{2,3,4}$.

\subsection{XOR algorithm with full homogeneous multivectors}

Any full homogeneous multivector $\mathbf{A}$ can be represented by 
\begin{equation}
 \mathbf{A} = \sum_{i=1}^{\binom{d}{g_a}} a_i \mathbf{e}_{\{ \mu_{i} \} } , 
\end{equation}
where $d$ is the dimension of the vector space, $g_a$ is the grade of $\mathbf{A}$. 

A straightforward solution to compute the number of operations required for the product of two full homogeneous multivectors $\mathbf{A}$ and $\mathbf{B}$ is to sum over the $\binom{d}{g_a}$ elements of the first multivector $\mathbf{A}$, combined to sum over the $\binom{d}{g_b}$ elements of the second multivector $\mathbf{B}$. Hereafter, this method will be referred as the double sum computation. 
Each product of this double sum computation will involve one addition/subtraction. Thus, the total number of required arithmetic operations for $p_{\wedge}^{\rm DS}$, $p_{\cdot}^{\rm DS}$, and $p_{*}^{\rm DS}$ each is
\begin{equation}
 p_{\wedge}^{\rm DS} = p_{ \cdot }^{\rm DS} = p_{*}^{\rm DS} =  2 \binom{d}{g_a} \binom{d}{g_b}.
    \label{eq:cplxtyDoubleSums}
\end{equation}

However, this double sum computation can be reduced. Indeed, the double sum does count even operations that lead to zero due to the nature of the product (e.g., in a $2$-dimensional vector space,  $(3 \mathbf{e}_1 + 4 \mathbf{e}_2) \wedge ( 5 \mathbf{e}_{12} ) = 0$). 
To the best of our knowledge, no previous work exists on the number of arithmetic operations that are really required in the outer, inner, and geometric product.

In the context of full homogeneous multivectors, the computational cost of looping over the two multivectors for the XOR algorithm is $\mathcal{O} ( \binom{d}{g_a} \binom{d}{g_b} ) $. As the computational cost of the sign computation is linear to the dimension, the complexity of the XOR algorithm is $\mathcal{O}(d \binom{d}{g_a} \binom{d}{g_b} )$. Note that this complexity is the same for the three products.

\subsection{Contributions} 
To address the lack of study on the number of arithmetic operations required for the outer, inner, and geometric products in practice, we give a study on the number of arithmetic operations for the outer, inner, and geometric products of geometric algebra for any full homogeneous multivectors, giving the tight number for the products by excluding operations that naturally lead to zero.  
This study is our base to prove that there exists an approach that reaches the equivalent complexity for each product.   
Table~\ref{tab:summarizeArithmeticOpe} summarizes the numbers of arithmetic operations required for full homogeneous multivectors.  Each of them is not greater than its corresponding number of arithmetic operations using the double sum computation. Table~\ref{tab:summarizeComplexity} gives the complexity of the products by the recursive approach over a prefix tree, which is strictly smaller than that of the XOR algorithm.  

\begin{table}[!ht]
    \centering
    \caption{
     Numbers of arithmetic operations required for products of two full homogeneous multivectors where $\mathcal{I} = \{|g_a - g_b |,|g_a - g_b |+2, \dots , g_a + g_b  \}$. Note that a full homogeneous multivector is a multivector with only one grade and all of the components of the grade are non-zero coefficients. 
    }
\label{tab:summarizeArithmeticOpe}
 \begin{tabular}{|c|c|} \hline 
 Outer product & $2\left(\begin{array}{@{}c@{}} d \\ g_a + g_b \end{array}\right)\left(\begin{array}{@{}c@{}} g_a+g_b \\ g_a \end{array}\right)$   \\
 \hline 
  Inner product & $ \displaystyle 2 \binom{d}{g_c} \binom{d-g_c}{ \displaystyle \frac{g_a+g_b-g_c}{2}} $  \\ \hline 
 Geometric product & $2\displaystyle\sum_{g_c \in \mathcal{I}}\left(\begin{array}{@{}c@{}} d \\ g_c \end{array}\right)\left(\begin{array}{@{}c@{}} g_c \\ \frac{ g_a - g_b + g_c}{2}\end{array}\right)\left(\begin{array}{@{}c@{}} d - g_c\\ \frac{g_a+g_b-g_c}{2} \end{array}\right)$ \\ \hline 
\end{tabular}
\renewcommand\arraystretch{1.0}
\end{table}

\begin{table}[!ht]
    \centering
    \caption{Complexity of the products by the recursive approach over a prefix tree .}
\label{tab:summarizeComplexity}
 \begin{tabular}{|c|c|} \hline 
Recursive outer product  & $ \displaystyle \mathcal{O} \Bigg( \binom{d}{g_a+g_b} \binom{g_a+g_b}{g_a} \Bigg) $  \\
 \hline 
 Recursive inner product & $ \displaystyle \mathcal{O}\Bigg( \binom{d}{g_c} \binom{d-g_c}{ \displaystyle \frac{g_a+g_b-g_c}{2}} \Bigg)$  \\
 \hline 
 Recursive geometric product  & $ \displaystyle \mathcal{O}\Bigg( \binom{d}{g_a} \binom{d}{g_b} \Bigg)$  \\
 \hline 
\end{tabular}
\end{table}

\section{Outer product}\label{sec:outer}

In implementations, the sign computation has a cost, i.e., the cost of counting the number of required permutations to have resulting basic vectors in the canonical order. In contrast, the number of arithmetic operations computed below corresponds to the number of operations found on a pre-computed code doing the product between two homogeneous multivectors. For this reason, the result given below omits the sign computation.

\subsection{Properties}
We denote by $\mathbf{A}\wedge\mathbf{B}$ the outer product between two homogeneous multivectors $\mathbf{A}$ and $\mathbf{B}$ with grades $g_a$ and $g_b$ in the $d$-dimensional vector space. This product $\mathbf{C}=\mathbf{A}\wedge\mathbf{B}$ can be defined from its property of distributivity over the addition:
\begin{equation}
\mathbf{C} = \sum_{k=1}^{\binom{d}{g_c}} c_k \mathbf{e}_{\{ \lambda_{k} \}} = \Bigg( \sum_{i=1}^{\binom{d}{g_a}} a_i \mathbf{e}_{\{ \mu_{i} \} } \Bigg) \wedge \Bigg(  \sum_{j=1}^{\binom{d}{g_b}} b_j \mathbf{e}_{\{ \nu_{j} \}}  \Bigg).
\label{eq:outerDoubleSum}
\end{equation}
As shown in~\cite{grassmann1844lineale}, the resulting multivector is homogeneous and its grade $g_c$ is
\begin{equation}
    g_c = g_a + g_b.
\end{equation}
Note that the grade of the resulting multivector has to be lower than or equal to the dimension of the vector space:
\begin{equation}
    g_a + g_b \leq d.
\end{equation}

\subsection{Number of arithmetic operations}


As stated in Section \ref{section:introduction}, there exist products that result in zero, even though their respective components are non-zero. In this section, we give a formula on the number of products that ignores such products that result in zero intrinsically. In practice, this number corresponds to the number of operations found on a pre-computed code doing the outer product between two homogeneous multivectors.  
\begin{thm}
Let $\mathbf{A}$ and $\mathbf{B}$ be homogeneous multivectors with grades $g_a$ and $g_b$. 
The number $p_{\wedge}^{\rm th}$ of the arithmetic operations involved in the outer product $\mathbf{C} = \mathbf{A}\wedge\mathbf{B}$ with grade $g_c$, where $g_c = g_a+g_b \leq d$ ($d$ is the dimension of the vector space) is given by
\begin{equation}
p_{\wedge}^{\rm th}=2\binom{d}{g_a + g_b} \binom{g_a + g_b}{ g_a} .
\label{eq:cplxtyRefinedOuterProd}
\end{equation}
\end{thm}

\begin{proof}
The outer product consists in splitting two basis blades into all possible basis blades of grade $g_a + g_b$ of the resulting multivector. This is equivalent to finding all the sub-blades whose grade is $g_a$ of the basis blades of $\mathbf{C}$. 
We know that there are 
$
    \binom{d}{g_a + g_b}
$
possible blades whose grade is $g_a + g_b$. 
On the other hand, the number of possible sub-blades of grade $g_a$ in any blade whose grade is $g_a + g_b$ is given as follows:
\begin{equation}
    \binom{g_a + g_b}{g_a}.
\end{equation}
Note that the above equation remains the same if we replace $g_a$ by $g_b$. This comes from the fact that by definition $g_b = g_c - g_a$. From the symmetry property of the binomial coefficient, we have
\begin{equation}
    \binom{g_a + g_b}{g_a} = \binom{g_a + g_b}{g_a + g_b - g_a} = \binom{g_a + g_b}{g_b}.
\end{equation}
Each product requires one addition. Hence, we obtain the total number of the required arithmetic operations by
\begin{equation}
    2 \binom{d}{g_a + g_b}\binom{g_a + g_b}{g_a}.
\end{equation}
\end{proof}

\subsection{Comparison with the double sum computation}
\label{sec:cmpDoublSumsWithRefinedOuter}
To see the difference between Equation~\eqref{eq:cplxtyRefinedOuterProd} and Equation~\eqref{eq:cplxtyDoubleSums},
let us compute the ratio between the two formulas as follows:
\begin{equation}
    \displaystyle \frac{p^{\rm th}_{\wedge}}{p^{\rm DS}_{\wedge}} = \frac{\displaystyle\binom{d}{g_a + g_b} \binom{g_a + g_b}{g_a} }{\displaystyle \binom{d}{g_a} \binom{d}{g_b}} .
\end{equation}
Using the trinomial revision property as defined in Chapter 5 of~\cite{mathsBinomial}, we have
\begin{equation}
    \displaystyle \frac{p^{\rm th}_{\wedge}}{p^{\rm DS}_{\wedge}} = \frac{\displaystyle\binom{d}{g_a} \binom{d-g_a}{g_b} }{\displaystyle \binom{d}{g_a} \binom{d}{g_b}} .
\end{equation}
After simplification $\Big(\forall 0 \leq g_a \leq d, \binom{d}{g_a} \neq 0\Big)$, we have
\begin{equation}
    \displaystyle \frac{p^{\rm th}_{\wedge}}{p^{\rm DS}_{\wedge}} = \frac{\displaystyle  \binom{d-g_a}{g_b} }{\displaystyle \binom{d}{g_b}} .
\end{equation}
The binomial coefficient $ \binom{n}{k} $ increases as $n$ increases when $k$ is fixed. Hence, this fraction is less than $1$. In practice, Equation~\eqref{eq:cplxtyRefinedOuterProd} may result in high improvements with respect to Equation~\eqref{eq:cplxtyDoubleSums}. As an example, let us assume that we compute the outer products of two trivectors in the algebra allowing to apply projective transformation of quadric surface, i.e. $8$-dimensional vector space~\cite{Goldman2015}. Then, Equation~\eqref{eq:cplxtyDoubleSums} requires approximately $5$ times more arithmetic operations than Equation~\eqref{eq:cplxtyRefinedOuterProd}: $560$ outer products instead of $3136$ in a $8$-dimensional vector space.

\section{Inner product}\label{sec:inner}

\subsection{Properties}
\label{sec:innerProductProperties}
The inner product $\mathbf{C}$ between two multivectors $\mathbf{A}$ and $\mathbf{B}$ with grades $g_a$ and $g_b$ is defined by
\begin{equation}
\mathbf{C} = \sum_{k=1}^{\binom{d}{g_c}} c_k \mathbf{e}_{\{ \lambda_{k} \}} = \Bigg( \sum_{i=1}^{\binom{d}{g_a}} a_i \mathbf{e}_{\{ \mu_{i} \} } \Bigg) \cdot \Bigg(  \sum_{j=1}^{\binom{d}{g_b}} b_j \mathbf{e}_{ \{ \nu_{j} \} }  \Bigg).
\label{eq:innerDoubleSum}
\end{equation}
This product is also distributive with respect to the addition. The resulting multivector is homogeneous and its grade is
\begin{equation}
    g_c = |g_a - g_b |.
\end{equation}
Note that when $g_a > g_b$, the product corresponds to the left contraction as defined in~\cite{LeoBook}. Whereas, when $g_b > g_a$, the resulting product is the right contraction. When $g_b=g_a$, on the other hand, the product becomes the scalar product.  

\subsection{Number of arithmetic operations}
\label{sec:innerprodRefined}
\begin{thm}
The number $p_{ \cdot }^{\rm th}$ of the arithmetic operations involved in the inner product $\mathbf{A}\cdot \mathbf{B}$ between two  homogeneous multivectors $\mathbf{A}$ and $\mathbf{B}$ with respective grades $g_a$ and $g_b$ is given by
\begin{equation}
\displaystyle p_{ \cdot }^{\rm th} =  2 \binom{d}{|g_a - g_b|} \binom{d-|g_a - g_b|}{\displaystyle  \frac{g_a+g_b- |g_a - g_b|}{2} } .
\label{eq:innerComplexity}
\end{equation}
\end{thm}

\begin{proof}
With the help of the set notation defined in Section~\ref{sec:notation}, the inner product between any two basis blades can be written as  
\begin{equation}
\mathbf{e}_{\lambda} = \mathbf{e}_{\mu} \cdot \mathbf{e}_{\nu} \qquad \lambda, \mu,\nu \in P(\mathcal{B}).
\end{equation}
By definition of the inner product in an orthogonal basis, we have two cases for $\mu$ and $\nu$ that lead to non-zero components.

The first case is
\begin{equation}
\mu \subseteq \nu. 
\end{equation}
In this case, $ | \lambda | = | \nu | - | \mu| $. By definition, the operation consists in the left contraction. Then
\begin{equation}
    \exists \beta,\gamma \in P(\mathcal{B}) \setminus \{ \varnothing  \}, \beta \cap \gamma = \varnothing , \mathbf{e}_{\lambda} = \mathbf{e}_{\beta} \cdot \mathbf{e}_{\beta \cup\gamma}.
\end{equation}
In such a case, $ g_c = |\lambda| = |\gamma | $ and $ \lambda = \gamma $. 
Computing the number of the products is reduced to determining the number of different possibilities for $\beta$ and $\gamma$. If we set $\gamma = \lambda$, then there is only one possibility for $\gamma$. 
As $ | \beta | + | \gamma | \leq d  \Rightarrow | \beta | \leq d - | \gamma | = d - g_c$, any possible grades of $\beta$ lower than or equal to $d-g_c$ is possible. As $ | \beta | = g_a$, any combination of $g_a$ in $d-g_c$ is possible.

In a similar way as for the outer product, any combination of $|\lambda| = g_c$ in $d$ is possible, which results in the number of the products as follows.
\begin{equation}
\label{eqn:first_configuration}
\binom{d}{g_c} \binom{d-g_c}{ g_a } .
\end{equation}
Furthermore, as $g_a \leq g_b $, $ |g_a - g_b| = g_b - g_a$. Then, Equation~\eqref{eqn:first_configuration} can be rewritten as
\begin{equation}
\begin{array}{cl}
\displaystyle \binom{d}{g_c} \binom{d-g_c}{\displaystyle \frac{2 g_a}{2} }      & = \displaystyle \binom{d}{g_c} \binom{d-g_c}{ \displaystyle \frac{g_a + g_b - g_b + g_a}{2} }  \\
     & = \displaystyle \binom{d}{g_c} \binom{d-g_c}{ \displaystyle \frac{g_a + g_b - | g_a - g_b | }{2} }.
\end{array}
\end{equation}

The second case is the symmetric case as follows:
\begin{equation}
 \nu \subseteq \mu    .
\end{equation}
Then,
\begin{equation}
    \exists \beta,\gamma \in P(\mathcal{B}) \setminus \{ \varnothing \}, \beta \cap \gamma = \varnothing, \mathbf{e}_{\lambda} = \mathbf{e}_{\beta \cup\gamma} \cdot \mathbf{e}_{\beta}.
\end{equation}
In this case, $ | \lambda | = | \mu | - | \nu| $. By definition, the operation results in the right contraction. 
Reasoning as in the previous paragraph leads us to $ g_c = |\lambda| = |\gamma | $ and $ \lambda = \gamma $. 
Computing the number of the products is then reduced to determining the number of different possibilities for $\beta$ and $\gamma$. If we set $\gamma = \lambda$, there is only one possibility for $\gamma$. 
As $ | \beta | + | \gamma | \leq d  \Rightarrow | \beta | \leq d - | \gamma | = d - g_c$. Therefore, any possible grade of $\beta$ lower than or equal to $d-g_c$ is possible. As $ | \beta | = g_b$, any combination of $g_b$ in $d-g_c$ is possible.

In a similar way as for the outer product, any combinations of $|\lambda| = g_c$ in $d$ is possible, resulting in the number of the products as follows.
\begin{equation}
\label{eqn:second_configuration}
\binom{d}{g_c} \binom{d-g_c}{ g_b } .
\end{equation}
Furthermore, as $g_b \leq g_a $, $ |g_a - g_b| = g_a - g_b$, Equation~\eqref{eqn:second_configuration} can be rewritten as
\begin{equation}
\begin{array}{cl}
\displaystyle \binom{d}{g_c} \binom{d-g_c}{ \displaystyle \frac{2 g_b}{2} }  &  = \displaystyle \binom{d}{g_c} \binom{d-g_c}{ \displaystyle\frac{g_a + g_b - g_a + g_b}{2} }  \\
     & = \displaystyle \binom{d}{g_c} \binom{d-g_c}{\displaystyle \frac{g_a + g_b - | g_a - g_b | }{2} }.
\end{array}
\end{equation}
Finally, since one product requires one addition, the total number of the required arithmetic operations is
\begin{equation}
2 \displaystyle \binom{d}{| g_a - g_b |} \binom{d-| g_a - g_b |}{\displaystyle \frac{g_a + g_b - | g_a - g_b | }{2} }.
\end{equation}
\end{proof}

\subsection{Comparison with the double sum computation}
\label{sec:cmpDoublSumsWithRefinedInner}
In a similar way as Section \ref{sec:cmpDoublSumsWithRefinedOuter}, let us compute the ratio between the two formulas \eqref{eq:innerComplexity} and \eqref{eq:cplxtyDoubleSums}:
\begin{equation}
\label{eqn:inner_ratio}
    \displaystyle \frac{p^{\rm th}_{\cdot}}{p^{\rm DS}_{\cdot}} = \frac{\displaystyle\binom{d}{| g_a - g_b |}  \binom{d-| g_a - g_b |}{\displaystyle \frac{g_a + g_b - | g_a - g_b | }{2} } }{\displaystyle \binom{d}{g_a} \binom{d}{g_b}} .
\end{equation}
If $g_a < g_b$, then Equation~\eqref{eqn:inner_ratio} can be rewritten as
\begin{equation}
    \displaystyle \frac{p^{\rm th}_{\cdot}}{p^{\rm DS}_{\cdot}} = \frac{\displaystyle\binom{d}{g_b - g_a}  \binom{d+ g_a - g_b }{g_a } }{\displaystyle \binom{d}{g_a} \binom{d}{g_b}} .
\end{equation}
Simplifying this equation can be achieved by revealing 
either $\binom{d}{g_b}$ or $\binom{d}{g_a}$ in 
its upper term.
This is merely performed through first applying the symmetry property of the binomial coefficient as follows.
\begin{equation}
\label{eqn:inner_ratio}
    \displaystyle \frac{p^{\rm th}_{\cdot}}{p^{\rm DS}_{\cdot}} = \frac{\displaystyle\binom{d}{d + g_a - g_b}  \binom{d+ g_a - g_b }{g_a} }{\displaystyle \binom{d}{g_a} \binom{d}{g_b}} .
\end{equation}
Then, Equation~\eqref{eqn:inner_ratio} can be simplified using the trinomial property defined in~\cite{mathsBinomial}:
\begin{equation}
\label{eqn:inner_simplified}
    \displaystyle \frac{p^{\rm th}_{\cdot}}{p^{\rm DS}_{\cdot}} = \frac{\displaystyle\binom{d}{g_a }  \binom{d - g_a }{d-g_b} }{\displaystyle \binom{d}{g_a} \binom{d}{g_b}} .
\end{equation}
For any grade and any dimension, $\binom{d}{g_a} \neq 0$. We thus simplify Equation~\eqref{eqn:inner_simplified} as below.
\begin{equation}
    \displaystyle \frac{p^{\rm th}_{\cdot}}{p^{\rm DS}_{\cdot}} = \frac{ \displaystyle \binom{d - g_a }{d-g_b} }{\displaystyle \binom{d}{g_b}} .
\end{equation}
Finally the symmetry property of the binomial coefficient applied to the left term yields
\begin{equation}
    \displaystyle \frac{p^{\rm th}_{\cdot}}{p^{\rm DS}_{\cdot}} = \frac{ \displaystyle \binom{d - g_a }{g_b-g_a} }{\displaystyle \binom{d}{g_b}}. 
\end{equation}
As for the outer product,  $\forall g_a \geq 0, \binom{d - g_a }{g_b-g_a} \leq \binom{d}{g_b} $. 

If $g_a \geq g_b$, a similar reasoning results in:
\begin{equation}
    \displaystyle \frac{p^{\rm th}_{\cdot}}{p^{\rm DS}_{\cdot}} = \frac{ \displaystyle \binom{d - g_b }{g_a-g_b} }{\displaystyle \binom{d}{g_a}} ,
\end{equation}
and the same conclusion holds.

Again Equation~\eqref{eq:innerComplexity} may result in high improvements with respect to Equation~\eqref{eq:cplxtyDoubleSums}. As an example, let us assume that we compute the inner products of two trivectors in a $8$-dimensional vector space. Then, Equation~\eqref{eq:cplxtyDoubleSums} requires $28$ times more arithmetic operations than Equation~\eqref{eq:innerComplexity}: $112$ arithmetic operations instead of $3136$ required to compute the inner product of two trivectors in a $8$-dimensional vector space.

\section{Geometric product}\label{sec:geometric}

\subsection{Properties}
We here deal with the geometric product. As mentioned in~\cite{perwassGeometric}, the possible grades of the resulting multivector are
\begin{equation}
    g_c \in \mathcal{I} = \{|g_a - g_b |,|g_a - g_b |+2, \dots , g_a + g_b  \}.
\end{equation}

The geometric product between two multivectors is then defined by
\begin{equation}
C = \sum_{ g_c \in \mathcal{I} } \smallskip \sum_{k=1}^{\binom{d}{g_c}}   c_k \mathbf{e}_{\{ \lambda_{k} \}} = \Bigg( \sum_{i=1}^{\binom{d}{g_a}} a_i \mathbf{e}_{\{ \mu_{i} \} } \Bigg) * \Bigg(  \sum_{j=1}^{\binom{d}{g_b}} b_j \mathbf{e}_{\{ \nu_{j} \} }  \Bigg).
\label{eq:standardGeoProduct}
\end{equation}
Note that in contrast to a multivector obtained by the outer product or the inner product, the resulting multivector might not be homogeneous. One might note that this contradicts the assumption that we deal with only full homogeneous multivectors. However, as stated in Section~\ref{section:introduction}, a non-homogeneous multivector is merely the sum of homogeneous multivectors. Moreover, the resulting homogeneous multivectors are also full. Thus, the assumptions still hold.


\subsection{Number of arithmetic operations}


\begin{thm}
The number $p^{\rm th}_{*}$ of the arithmetic operations involved in the geometric product $\mathbf{A} * \mathbf{B}$ between two  homogeneous multivectors $\mathbf{A}$ and $\mathbf{B}$ with respective grades $g_a$ and $g_b$ is given by
\begin{equation} 
\displaystyle p^{\rm th}_{*} = 2 \sum_{g_c \in \mathcal{I}}  \binom{d}{g_c} \binom{g_c}{ \displaystyle \frac{ g_a - g_b + g_c }{2} } \binom{d-g_c}{\displaystyle \frac{g_a + g_b - g_c}{2}},
\label{eq:numberProductsGeoProduct}
\end{equation}
where $\mathcal{I} = \{|g_a - g_b |,|g_a - g_b |+2, \dots , g_a + g_b  \}$.
\end{thm}

\begin{proof}
The geometric product between any two basis blades can be written as  
\begin{equation}
\label{eqn:geometric_product_with_grades}
\mathbf{e}_{\lambda} = \mathbf{e}_{\mu} * \mathbf{e}_{\nu} \qquad  \lambda, \mu,\nu \in P(\mathcal{B}).
\end{equation}
There are four cases with respect to $\mu$ and $\nu$. 

The first case is
\begin{equation}
\mu \cap \nu = \varnothing .     
\end{equation}
Then, the geometric product results in the outer product between the basis blades, and the number of products is already shown.

The second case corresponds to
\begin{equation}
\mu \subseteq \nu .    
\end{equation}
In this case, $ | \lambda | = | \nu | - | \mu| $. By definition, the operation results in the left contraction. The number of the products is addressed in Section~\ref{sec:innerprodRefined}.

The third case corresponds to
\begin{equation}
 \nu \subseteq \mu . 
\end{equation}
In this case, $ | \lambda | = | \mu | - | \nu| $. By definition, the operation is reduced to the right contraction. The computation of the number of products is already addressed in Section~\ref{sec:innerprodRefined}.

Finally, the last case is the situation where $\mu \cap \nu \neq \emptyset $ but $\nu \nsubseteq \mu$ nor $\mu \nsubseteq \nu$. More precisely, this corresponds to
\begin{equation}
    \exists \alpha,\beta,\gamma \in P(\mathcal{B}) \setminus \{ \varnothing \}, \alpha \cap \beta = \varnothing, \beta \cap \gamma = \varnothing, \mathbf{e}_{\lambda} = \mathbf{e}_{\alpha \cup\beta} * \mathbf{e}_{\beta \cup\gamma}.
\end{equation}
In such a case, $ g_c = |\lambda| = |\alpha | + | \gamma | $ and $ \lambda = \alpha\cup \gamma $. 
Computing the number of products is reduced to determining the number of different possibilities for $\alpha,\beta$, and $\gamma$. Let us start with $\beta$.
The union of the two blades of Equation~\eqref{eqn:geometric_product_with_grades} results in
\begin{equation}
    \mathbf{e}_{\lambda} = \mathbf{e}_{\alpha \cup \beta \cup \beta \cup \gamma} .
\end{equation}
Therefore, 
\begin{equation}
\begin{array}{ccl}
    |  \alpha \beta | + |  \beta \gamma | - | \beta \beta |   &=& |  \alpha \beta | + |  \beta \gamma | - 2| \beta |, \\
    g_c &=& g_a + g_b - 2 | \beta |.
\end{array}    
\end{equation}
Hence, we have
\begin{equation}
    | \beta | = \frac{ g_a + g_b -g_c }{2}.
    \label{eq:cardBeta}
\end{equation}
Due to the fact that $\beta \cap \gamma = \varnothing$ and $\beta \cap \alpha = \varnothing$, $\beta \cap \lambda = \varnothing$. Thus,
\begin{equation}
    \beta \in P(\mathcal{B}) \setminus \{ \lambda,\varnothing \}, | \beta | = \frac{ g_a + g_b -g_c }{2} .
\end{equation}
Furthermore, 
\begin{equation}
    \beta \in P(\mathcal{B})\setminus P(\lambda). 
\end{equation}
Since the set of maximal cardinality in $P(\mathcal{B})\setminus  P(\lambda)$ is $d - g_c$,
the number of possibilities for $\beta$ is the number of possible combinations of $| \beta |$ in   $d - g_c$:
\begin{equation}
    \binom{d - g_c}{\displaystyle \frac{ g_a + g_b -g_c }{2}} . 
\end{equation}
Note that we have to ensure that $\frac{ g_a + g_b -g_c }{2}$ is an integer. Two cases exist: either $g_a$ and $g_b$ have the same parity or not. 

If both $g_a$ and $g_b$ have the same parity, then
\begin{equation}
    \begin{array}{cl}
   	\exists n \in \mathbb{Z}, g_a + g_b &= 2 n, \\
   	\exists n' \in \mathbb{Z}, \lvert g_a - g_b \rvert  &= 2 n'.
    \end{array}
\end{equation}
Furthermore, $g_c$ is the sum of $\lvert g_a - g_b \rvert$ and an even number, thus $g_c$ is also even. Since the sum of two even numbers is also even, $g_a + g_b - g_c$ is even.

Now let assume that $g_a$ and $g_b$ do not have the same parity. Then, their sum and their difference are both odds. On the other hand, $g_c$ is the sum of $\lvert g_a - g_b \rvert$ and an even number, indicating that the $g_c$ is odd. Since the difference of two odd numbers is even, $g_a + g_b - g_c$ is even. Hence, in both cases, $g_a + g_b - g_c$ is even.

The number of the combinations for $\alpha$ and $\gamma$ is now computed. We know that $g_c = |\lambda| = |\alpha | + | \gamma |$ and $\lambda = \alpha \gamma$. Thus, the number of the combinations in this case is merely equivalent to the number of possibilities of the outer product associated to $g_c$ and $\alpha,\gamma$: 
\begin{equation}
    \binom{g_c}{|\alpha|} = \binom{g_c}{|\gamma|} .
\end{equation}
Furthermore, $g_a = |\alpha| + |\beta| \Rightarrow |\alpha | = g_a - |\beta |$ and using the definition of $|\beta | $ in Equation~\eqref{eq:cardBeta} results in a number of possibilities
\begin{equation}
\renewcommand\arraystretch{2.2}
\begin{array}{cl}
     \displaystyle \binom{g_c}{|\alpha|} &= \displaystyle \binom{g_c}{ g_a - |\beta | }  \\
                           &= \displaystyle \binom{g_c}{ g_a - \displaystyle \frac{ g_a + g_b -g_c }{2} }  \\
                           &= \displaystyle \binom{g_c}{ \displaystyle \frac{ g_a - g_b + g_c }{2} }.
\end{array}
\renewcommand\arraystretch{1.0}
\end{equation}

Note that for the same reason as in the above paragraphs, the term $g_a - g_b + g_c$ is even. Furthermore $g_a - g_b + g_c \geq 0$ because by assumption $g_c > |g_a -g_b| $.
This results in a number of products of
\begin{equation}
\label{eqn:Geometric_simplified}
    \binom{d}{g_c} \binom{g_c}{ \displaystyle \frac{ g_a - g_b + g_c }{2} } \binom{d-g_c}{\displaystyle \frac{g_a + g_b - g_c}{2}}.
\end{equation}
Equation~\eqref{eq:standardGeoProduct} shows that the geometric product is the sum over all possible grades $g_c \in \mathcal{I}$. Accordingly, Equation~\eqref{eqn:Geometric_simplified} yields
\begin{equation}
    \sum_{g_c \in \mathcal{I}} \binom{d}{g_c} \binom{g_c}{ \displaystyle \frac{ g_a - g_b + g_c }{2} } \binom{d-g_c}{\displaystyle \frac{g_a + g_b - g_c}{2}}.
\end{equation}
Finally, as each product requires one addition, the total number of the arithmetic operations is
\begin{equation}
        2 \sum_{g_c \in \mathcal{I}} \binom{d}{g_c} \binom{g_c}{ \displaystyle \frac{ g_a - g_b + g_c }{2} } \binom{d-g_c}{\displaystyle \frac{g_a + g_b - g_c}{2}}.
\end{equation}
\end{proof}

\subsection{Comparison with the double sum computation}
\begin{prop}
\begin{equation}
\label{eqn:geometric_product_eq}
        \sum_{g_c \in \mathcal{I}} \binom{d}{g_c} \binom{g_c}{ \displaystyle \frac{ g_a - g_b + g_c }{2} } \binom{d-g_c}{\displaystyle \frac{g_a + g_b - g_c}{2}} = \binom{d}{g_a} \binom{d}{g_b}.
\end{equation}
\label{proposition:geometricProduct}
\end{prop}
\begin{proof}

In addition to the symmetry property and the trinomial property, we will use here the Vandermonde's convolution property of the binomial coefficient whose proof can be found in Chapter 5 of~\cite{mathsBinomial}. 
We first introduce a variable to drop divisions. Let us define 
\begin{equation}
s=\frac{g_b - g_a + g_c }{2}.
\end{equation}
Let us assume, without loss of generality, that $g_a > g_b$.   
Then, as $g_c \in \mathcal{I} = \{|g_a - g_b |,|g_a - g_b |+2, \dots , g_a + g_b  \} $, 
\begin{equation}
 s \in \{ 0, 1, \cdots, g_b \}.   
\end{equation}
This yields
\begin{equation}
\label{eqn:Geometric_modified}
\begin{array}{rl}
    & \displaystyle \sum_{g_c \in \mathcal{I}} \binom{d}{g_c} \displaystyle \binom{g_c}{ \displaystyle \frac{ g_a - g_b + g_c }{2} } \binom{d-g_c}{\displaystyle \frac{g_a + g_b - g_c}{2}} \\
    =& \displaystyle \sum_{s=0}^{g_b} \binom{d}{2s + g_a - g_b } \binom{2s + g_a - g_b}{  s + g_a - g_b  } \binom{d- 2s + g_b -g_a}{g_b - s}.
\end{array}
\end{equation}
We apply the trinomial revision property to the two left-most terms in Equation~\eqref{eqn:Geometric_modified}, resulting in
\begin{equation*}
\begin{array}{rl}
    & \displaystyle \sum_{s=0}^{g_b} \binom{d}{s+g_a-g_b } \binom{d-s + g_b - g_a}{  s  } \binom{d- 2s + g_b -g_a}{g_b - s}.
\end{array}
\end{equation*}
Next, we apply the same property to the two right-most terms, yielding
\begin{equation}
\begin{array}{rl}
    & \displaystyle \sum_{s=0}^{g_b} \binom{d}{s+g_a-g_b} \binom{d-s + g_b - g_a}{  g_b  } \binom{g_b}{ s}.
\end{array}
\end{equation}
The symmetry property is then applied to the left-most term.  We have
\begin{equation}
\label{eqn:Geometric_simple_modified}
\begin{array}{rl}
    & \displaystyle \sum_{s=0}^{g_b} \binom{d}{d-s+g_b-g_a} \binom{d-s + g_b - g_a}{  g_b  } \binom{g_b}{ s}.
\end{array}
\end{equation}
Again, we apply the trinomial revision property 
to the two left-most terms in Equation~\eqref{eqn:Geometric_simple_modified}.  We now have  
\begin{equation}
\begin{array}{rl}
\label{eqn:simple}
    & \displaystyle \binom{d}{g_b} \sum_{s=0}^{g_b} \binom{d - g_b}{ d-s-g_a } \binom{g_b}{s}.
\end{array}
\end{equation}
Note that $\binom{d}{g_b}$ does not depend on $s$.
Applying 
the Vandermonde's convolution property
to Equation~\eqref{eqn:simple} results in
\begin{equation}
\begin{array}{rl}
    & \displaystyle \binom{d}{g_b} \binom{d }{ d-g_a }.
\end{array}
\end{equation}
After using the symmetry property on the right term, we see Equation~\eqref{eqn:geometric_product_eq} holds.  
\end{proof}

\section{Generating products by the recursive approach over a prefix tree} 
\label{sec:recursiveApproach}

We show here that the prefix tree algorithm used in Garamon~\cite{breuils_garamon_2019} reaches the above derived numbers of arithmetic operations for products between two full homogeneous multivectors.  

To make the paper self-contained, let us briefly review in Section~\ref{section:prefix-tree} the recursive formulation~\cite{breuils_garamon_2019} to define multivectors and geometric algebra products (see \cite{breuils_garamon_2019} for more details).
We start with the definition of multivectors using the prefix tree structure.


\subsection{Multivectors}
\label{section:prefix-tree}
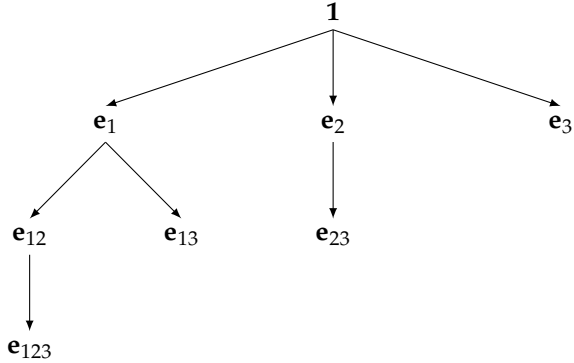
\begin{figure}[!ht]
 \begin{tikzpicture}
 	\tikzset{primal/.style={->,>=latex,thin}}
 	\node         (1) at (0,4.5) {$\mathbf{1}$};  
	\node         (e1) at (-3,3) {$\mathbf{e}_{1}$};  
 	\node         (e2) at (0,3) {$\mathbf{e}_{2}$};  
 	\node         (e3) at (3,3) {$\mathbf{e}_{3}$};  
 	\node         (e12) at (-4,1.5) {$\mathbf{e}_{12}$};  
 	\node         (e13) at (-2,1.5) {$\mathbf{e}_{13}$};  
 	\node         (e23) at (0,1.5) {$\mathbf{e}_{23}$};  
 	\node         (e123) at (-4,0) {$\mathbf{e}_{123}$};  
 	\draw[primal] (1.south) -- (e1.north);
 	\draw[primal] (1.south) -- (e2.north);
 	\draw[primal] (1.south) -- (e3.north);
 	\draw[primal] (e1.south) -- (e12.north);
 	\draw[primal] (e1.south) -- (e13.north);
   \draw[primal] (e2.south) -- (e23.north);
   \draw[primal] (e12.south) -- (e123.north);
 \end{tikzpicture}\\
 \caption{Prefix tree structure of the basis blades for a geometric algebra whose underlying vector space is of dimension~$3$.
 }
 \label{fig::prefixtree}
 \end{figure}

Each basis blade is associated with a node of a prefix tree and the nodes of depth $k$ in the prefix tree correspond to the basis blades of grade $k$. 
Thus, the scalar basis blade, denoted by $\mathbf{1}$, is associated with the root node. The vector basis blades are associated with the children of the root node, the bivector basis blades are associated to the children of those nodes, and so on, as illustrated on Figure~\ref{fig::prefixtree}.
By construction of the prefix tree, the index of a basis blade associated with a node is prefixed by the indexes of the basis blades associated with its parent nodes.
Note that the breadth-first search of the basis blades over the prefix tree results in the list of basis blades in the canonical order. 
For instance, the list obtained from the prefix tree in Figure~\ref{fig::prefixtree} is
$(\mathbf{1},\mathbf{e}_{1},\mathbf{e}_{2},\mathbf{e}_{3},\mathbf{e}_{12},\mathbf{e}_{13},\mathbf{e}_{23},\mathbf{e}_{123})$.

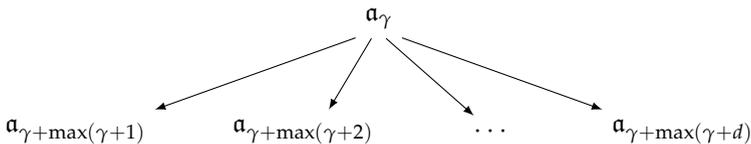
\begin{figure}[ht!]
\centering
\begin{tikzpicture}
	\tikzset{primal/.style={->,>=latex,thin}}
	\node         (root) at (0,4.5) {$\mathfrak{a}_{\gamma}$};  
	\node         (e1) at (-4,3) {$\mathfrak{a}_{\gamma + \max(\gamma + 1)}$};  
	\node         (e2b) at (1.5,3) { $\cdots$}; 
	\node         (e2) at (-1,3) {$\mathfrak{a}_{\gamma + \max(\gamma + 2)}$}; 
	\node         (e3) at (4,3) {$\mathfrak{a}_{\gamma + \max(\gamma + d)}$};  
	\draw[primal] (root.220) -- (e1.north east);
	\draw[primal] (root.250) -- (e2.40);
	\draw[primal] (root.290) -- (e2b.140);
	\draw[primal] (root.320) -- (e3.north west);
\end{tikzpicture}
\caption{Labelling of the siblings of a child node.}
\label{fig:labellingInformation}
\end{figure}

Given a multivector $\mathbf{A}$, let us assume that $\mathfrak{a}_{\gamma}$ represents a node of the prefix tree, where $\gamma$ is the set of basis vectors present in the basis blade. For example, the node $\mathfrak{a}_{\gamma} = \mathfrak{a}_{\{1,2\}}$ corresponds to the node associated with the blade~$\mathbf{e}_{12}$ of~$\mathbf{A}$. Then, the set of  children of any node $\mathfrak{a}_{\gamma}$ can be recursively defined from depth $n$ to the next depth $n+1$ as follows:
\begin{equation}
\begin{array}{r@{}lcl}
\text{a node at depth:~} & n   & \rightarrow & \mathfrak{a}_{\gamma}, \\
\text{its children at depth:~} & n+1 & \rightarrow & \mathfrak{a}_{\gamma + \mu} \text{, } ~~~ \mu \in [ \max( \gamma )+1, \cdots, d ],
\end{array}
\label{eq:PrefixTreeCompactNot}
\end{equation}
where $d$ is the dimension of the vector space.
Note that the function $\max()$ is self-sufficient since the integer is a totally ordered set. Furthermore, the addition sign between two sets (greek letters) denotes the concatenation of the two sets. An illustration of the recursion from a node to its children is given in Figure~\ref{fig:labellingInformation}. The starting call of the recursive formula for the breadth-first search is $\mathfrak{a}_{0}$ at a depth of $0$ (grade~$0$ or scalar). The end of recursion is achieved when a node is a leaf (i.e. $\max(\gamma) = d$).

\subsection{Recursive outer product}
The recursive outer product was introduced by~\cite{fuchsThery2014} and defined over the binary tree in~\cite{Breuils2017}, then was adapted for the prefix tree in~\cite{breuils_garamon_2019}. The resulting complexity of this recursive method for full multivectors in $d$-dimensional space is $\mathcal{O}(3^d)$. This section aims at computing the complexity for full \textit{homogeneous} multivectors. 
We first remind the recursive outer product for general multivectors and then update this product for homogeneous multivectors.
\begin{defn}[Recursive outer product over a prefix tree for general multivectors]
\label{def:recursive_outer_product_general}
Given two general multivectors $\mathbf{A}$ and $\mathbf{B}$, the recursive outer product associated with $\mathbf{C} = \mathbf{A} \wedge \mathbf{B}$ is expressed as
\begin{equation}
\begin{array}{r@{}l}
\text{at depth } n \quad\\
\text{computation:~}   & ~~\mathfrak{c}_{\lambda}  \mathrel{+}= \mathfrak{a}_{\gamma} \wedge \mathfrak{b}_{\delta} \\ 
\text{recursive calls:~}  & \mathfrak{c}_{\lambda + \sigma} = \mathfrak{a}_{\gamma+ \sigma} \wedge \mathfrak{b}_{\delta} + \overline{\mathfrak{a}_{\gamma}} \wedge \mathfrak{b}_{\delta+ \sigma} \text{, } ~~~ \sigma \in [ \max( \lambda )+1, \cdots, d ] 
\end{array},
\label{eq:recursiveFormulaPrefixOuterproductGeneral}
\end{equation} 
where the overline denotes the anticommutativity property of the product.
\end{defn}
The starting call of the recursive formula is $\mathfrak{c}_{0} = \mathfrak{a}_{0} \wedge \mathfrak{b}_{0}$, i.e. at a depth of $0$ (grade $0$ or scalar). The end of recursion is achieved when a node is a leaf (i.e. $\max(\lambda) = d$). 
\begin{defn}[Anticommutativity]
The recursive construction of the anticommutativity of multivector $\mathbf{A}$ is
\begin{equation}
\begin{array}{ccl}
n   & \rightarrow & \phantom{-} \overline{\mathfrak{a}_{\gamma}}  \\
n+1 & \rightarrow & -\overline{\mathfrak{a}_{\gamma+\mu}} \text{, } \quad \mu \in [ \max( \gamma )+1, \cdots, d ] 
\end{array}.
\label{eq:recursiveFormulaPrefixAnticommutativity}
\end{equation} 
\end{defn}

In the case of homogeneous multivectors, the grades $g_a$, $g_b$, and $g_c$ of the multivectors are known in advance. The recursive product $\mathbf{C} = \mathbf{A} \wedge \mathbf{B}$ can then be slightly modified so that any update of $\mathfrak{c}$ are performed only at depth $g_c = g_a+g_b$. 
\begin{defn}[Recursive outer product of homogeneous multivectors over a prefix tree]
\label{def:recursive_outer_product}
Given two full homogeneous multivectors $\mathbf{A}$ and $\mathbf{B}$ of respective grade $g_a$ and $g_b$, the recursive outer product associated with $\mathbf{C} = \mathbf{A} \wedge \mathbf{B}$ of expected grade $g_c$ is expressed as
\begin{equation}
\begin{array}{r@{}l}
\text{at depth } n \quad\\
\text{computation:~}   & ~~\mathfrak{c}_{\lambda}  \mathrel{+}= \mathfrak{a}_{\gamma} \wedge \mathfrak{b}_{\delta},\quad\text{if }|\lambda| = g_c  \\ 
\text{recursive calls:~}  & \mathfrak{c}_{\lambda + \sigma} = \mathfrak{a}_{\gamma+ \sigma} \wedge \mathfrak{b}_{\delta} + \overline{\mathfrak{a}_{\gamma}} \wedge \mathfrak{b}_{\delta+ \sigma} \text{, } ~~~ \sigma \in [ \max( \lambda )+1, \cdots, d ] 
\end{array},
\label{eq:recursiveFormulaPrefixOuterproduct}
\end{equation} 
where $|\lambda|$ denotes the cardinality of the set $\lambda$.
\end{defn}
Thus, for homogeneous multivectors, the end of recursion is achieved when a node is a leaf (i.e. $\max(\lambda) = d$) or when the targeted grade $g_c$ is reached (i.e. $|\lambda| = g_c$). 
Algorithm~\ref{algo:recursiveOuterProduct} presents a straightforward way to implement the recursive formulas presented in Definitions~\ref{def:recursive_outer_product_general} and~\ref{def:recursive_outer_product}. 
\begin{algorithm}[!ht]
\DontPrintSemicolon
\Fn{ \funct{outer}}{
\KwIn{~$\mathfrak{a}_{\gamma}, \mathfrak{b}_{\delta}$: node of multivectors $\mathbf{A}$ and $\mathbf{B}$,\\
{\Indp \Indp $\mathfrak{c}_{\lambda}$: nodes of the resulting multivector $\mathbf{C}$\\}
{\Indp \Indp $\mathtt{complement}$: recursive value ( $\pm 1$).\\}
{\Indp \Indp $\mathtt{sign}$: recursive sign coefficient ( $\pm 1$).\\}
}
\SetKw{and}{and}
\SetKw{In}{in}
\SetKw{Suchthat}{such that}
{
\If({\textcolor{gray}{$\quad\mathbin{/\mkern-6mu/}$ remove this condition for general multivectors}} \label{alg:recursiveOuterProduct_grade_optimization}){$\vert\lambda\vert = g_c$}{$\mathfrak{c}_{\lambda} \mathrel{+}= \mathtt{sign} \times \mathfrak{a}_{\gamma} \times \mathfrak{b}_{\delta}  $ \;} 
\ForEach{  $\sigma \in [ \max( \lambda )+1, \cdots, d ]$  } {   
 \textcolor{gray}{// $\mathfrak{a}_{\gamma+ \sigma} \wedge \mathfrak{b}_{\delta}$}\; 
    \funct{outer}($\mathfrak{a}_{\gamma+ \sigma},\mathfrak{b}_{\delta}, \mathfrak{c}_{\lambda+\sigma},\mathtt{sign} \times \mathtt{complement},\mathtt{complement}$) \; 
    \textcolor{gray}{// $\overline{\mathfrak{a}_{\gamma}} \wedge \mathfrak{b}_{\delta+ \sigma}$}\;
    \funct{outer}($\mathfrak{a}_{\gamma},\mathfrak{b}_{\delta+\sigma},\mathfrak{c}_{\lambda+\sigma},\mathtt{sign},-\mathtt{complement} $)  \;
}
}
}
First call: \funct{outer}( $\mathfrak{a}_{0},\mathfrak{b}_{0},\mathfrak{c}_{0},1,1$)
\caption{Pseudo-code of the recursive outer product $\mathbf{C} = \mathbf{A} \wedge \mathbf{B}$ }
\label{algo:recursiveOuterProduct}
\end{algorithm}

 \begin{figure}[!ht]
 \begin{tikzpicture}
 	\tikzset{primal/.style={->,>=latex,thin}}
 	\node         (1) at (0,4.5) {$\mathfrak{a}_0 \wedge \mathfrak{b}_0$};  
	\node         (e1) at (-3,3) {$\mathfrak{a}_{1} \wedge \mathfrak{b}_{0}+ \mathfrak{a}_0 \wedge \mathfrak{b}_1$};  
 	\node         (e2) at (0,3) {$\mathfrak{a}_{2} \wedge \mathfrak{b}_{0}+ \mathfrak{a}_0 \wedge \mathfrak{b}_2$};  
 	\node         (e3) at (3,3) {$\mathfrak{a}_{3} \wedge \mathfrak{b}_{0}+ \mathfrak{a}_0 \wedge \mathfrak{b}_3$};  
 	\node         (e12) at (-4,1) {$ 
 	\begin{array}{c}
 	     \mathfrak{a}_{12} \wedge \mathfrak{b}_{0} \\
 	       -\mathfrak{a}_{2} \wedge \mathfrak{b}_{1} \\
 	       +\mathfrak{a}_{1} \wedge \mathfrak{b}_{2} \\
           -\mathfrak{a}_{0} \wedge \mathfrak{b}_{12} \\
    \end{array}$
    };  
 	\node         (e13) at (-2,1) {$ 
 	\begin{array}{c}
 	     \mathfrak{a}_{13} \wedge \mathfrak{b}_{0} \\
 	       -\mathfrak{a}_{3} \wedge \mathfrak{b}_{1} \\
 	       +\mathfrak{a}_{1} \wedge \mathfrak{b}_{3} \\
           -\mathfrak{a}_{0} \wedge \mathfrak{b}_{13} \\
    \end{array}$
    };    
 	\node         (e23) at (0,1) {$ 
 	\begin{array}{c}
 	     \mathfrak{a}_{23} \wedge \mathfrak{b}_{0} \\
 	       -\mathfrak{a}_{3} \wedge \mathfrak{b}_{2} \\
 	       +\mathfrak{a}_{2} \wedge \mathfrak{b}_{3} \\
           -\mathfrak{a}_{0} \wedge \mathfrak{b}_{23} \\
    \end{array}$
    };    
 	\node         (e123) at (-4,-2.5) {$ 
 	\begin{array}{c}
 	     \mathfrak{a}_{123} \wedge \mathfrak{b}_{0} \\
 	       +\mathfrak{a}_{12} \wedge \mathfrak{b}_{3} \\
 	       +\mathfrak{a}_{23} \wedge \mathfrak{b}_{1} \\
           -\mathfrak{a}_{2} \wedge \mathfrak{b}_{13} \\
   	       -\mathfrak{a}_{13} \wedge \mathfrak{b}_{2} \\
 	       +\mathfrak{a}_{1} \wedge \mathfrak{b}_{23} \\
 	       +\mathfrak{a}_{3} \wedge \mathfrak{b}_{12} \\
           +\mathfrak{a}_{0} \wedge \mathfrak{b}_{123} \\
    \end{array}$
    };  
 	\draw[primal] (1.south) -- (e1.north);
 	\draw[primal] (1.south) -- (e2.north);
 	\draw[primal] (1.south) -- (e3.north);
 	\draw[primal] (e1.south) -- (e12.north);
 	\draw[primal] (e1.south) -- (e13.north);
   \draw[primal] (e2.south) -- (e23.north);
   \draw[primal] (e12.south) -- (e123.north);
 \end{tikzpicture}\\
 \caption{Prefix tree structure associated with the recursive outer product for a geometric algebra whose underlying vector space is of dimension~$3$. Note that for a given depth, each node presents the same number of outer products.
 }
 \label{fig::prefixtreeOuterProductDim3}
 \end{figure}
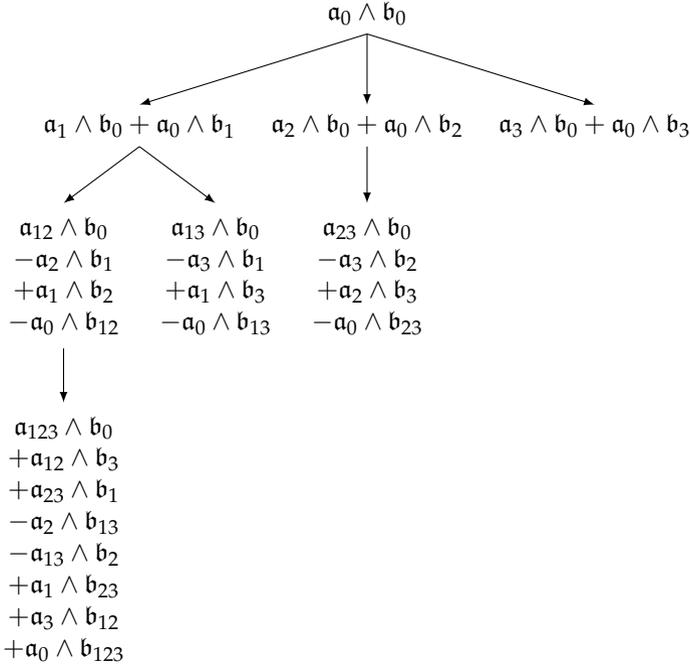
 
Figure~\ref{fig::prefixtreeOuterProductDim3} illustrates an example of the development of all the recursive outer products in the $3$-dimensional vector space. The number of recursive calls depends only on the depth of the recursion, as stated with the following lemma whose proof is given in Appendix~\ref{proof:sameOuterProductsEachNode}.

\begin{lem}
During the recursive product $\mathbf{C}=\mathbf{A}\wedge\mathbf{B}$, all the children $\mathfrak{c}_{\lambda+\sigma}$ of a node $\mathfrak{c}_{\lambda}$ of the the prefix tree corresponding to $\mathbf{C}$ generate the same number of recursive outer products calls. 
In other words, the siblings at any depth of the prefix tree of $\mathbf{C}$ generate the same number of products.
\label{lem:sameProductsForAllBrothers}
\end{lem}

\begin{thm}
The complexity $c_\wedge^{\rm rec}$ of the recursive outer product $\mathbf{C} = \mathbf{A} \wedge \mathbf{B}$ between two homogeneous multivectors $\mathbf{A}$ and $\mathbf{B}$  of respective grade $g_a$ and~$g_b$, with resulting grade~$g_c = g_a+g_b$, is expressed as
\begin{equation}
c_\wedge^{\rm rec} 
= \mathcal{O}  \Bigg( \binom{d}{g_a + g_b} \binom{g_a + g_b}{ g_a} \Bigg),
\label{eq:cplxtyRecOuterProd}
\end{equation}
where $d$ is the dimension of the vector space.
\end{thm}

\begin{proof}
Lemma~\ref{lem:sameProductsForAllBrothers} shows that during a recursive outer product, the siblings at any depth (grade) of the resulting prefix tree have the same number of outer products, i.e, the same number of recursive calls.
Furthermore, there are $\binom{d}{g_c}$ nodes of grade $g_c$ in the prefix tree represented in the $d$-dimensional vector space. The number of products of a given depth is thus the multiplication of the binomial coefficient and $n_{g_c,g_a}$ (the number of outer products per node of grade~$g_c$). Hence, the overall complexity is
\begin{equation}
    c_\wedge^{\rm rec} = \mathcal{O} \Bigg( \binom{d}{g_c} n_{g_c,g_a} \Bigg).
 \label{eq:outerprodComplexity_step1}
\end{equation}

We may focus on 
the computation of $n_{g_c,g_a}$, accordingly.
%
The recursive formula of Definition~\ref{def:recursive_outer_product}
shows that at any depth of recursion, there is a sum of two recursive calls to be executed. Both the recursive calls increase the grade of the result. One increases the grade of $\mathfrak{a}$ and the other leaves it unchanged. 
Applying the recursion in the forward order yields
\begin{equation}
    \begin{array}{ccc}
        n_{0,0} & = & n_{1,1} + n_{1,0}, \\
        n_{1,1} & = & n_{2,2} + n_{2,1}, \\
        n_{1,0} & = & n_{2,1} + n_{2,0}, \\
                & \vdots & \\
        n_{g_c-2,g_a-1} & = & n_{g_c-1,g_a} + n_{g_c-1,g_a-1}, \\
        n_{g_c-1,g_a} & = & n_{g_c,g_a}, \\
        n_{g_c-1,g_a-1} & = & n_{g_c,g_a}.
    \end{array}
\end{equation}
When the final recursion is reached for the grade of $g_c$, two recursive calls $n_{g_c,g_a-1}$ and $n_{g_c,g_a+1}$ (corresponding to the respective ending conditions of the two terms of Equation~\eqref{eq:recursiveFormulaPrefixOuterproduct}) are not executed. Now, going backward from the two final recursion equations $n_{g_c,g_a}$ yields the recursive formula
\begin{equation}
    \begin{array}{ccc}
        n_{g_c,g_a} & = & n_{g_c-1,g_a} + n_{g_c-1,g_a-1}.
    \end{array}
\end{equation}
We verify that the cases where either $g_c = g_a$ or $g_a = 0$ correspond to a final recursion condition, and, thus, we have $n_{g_c,g_a} = 1$. 
This recursive definition corresponds to the recursive definition of the binomial coefficient:
\begin{equation}
    \binom{g_c}{g_a} = \binom{g_c-1}{g_a} + \binom{g_c-1}{g_a-1}.
\end{equation}
Hence, the number of recursive calls is thus
\begin{equation}
n_{g_c,g_a} = \binom{g_c}{g_a} = \binom{g_a + g_b}{g_a}.
\end{equation}
The complexity of the recursive outer product is 
\begin{equation}
c_\wedge^{\rm rec} = \mathcal{O} \Bigg(  \binom{d}{g_c} n_{g_c,g_a} \Bigg) = 
\mathcal{O} \Bigg( \binom{d}{g_a + g_b} \binom{g_a + g_b}{g_a} \Bigg). 
\end{equation}
\end{proof}

\begin{figure}[H]
\centering
\begin{subfigure}{0.85\textwidth}
\begin{tikzpicture}[scale=0.60, every node/.style={scale=0.85}]
	\tikzset{primal/.style={->,>=latex,thin}}
	\tikzset{useful/.style={->,>=latex,thin,color=black}}
	\tikzset{useless/.style={dashed,->,>=latex,thin,color=blue}}
	\tikzset{uselessover/.style={dashed,->,>=latex,thin,color=forestGreen}}
	\node         (1) at (0,6) {$\mathbf{1}$};  
	\node         (e1) at (-5,4.5) {$\mathbf{e}_{1}$};  
	\node         (e2) at (-1,4.5) {$\textcolor{blue}{\mathbf{e}_{2}}$};  
	\node         (e3) at (2,4.5) {$\textcolor{blue}{\textcolor{blue}{\mathbf{e}_{3}}}$};  
	\node         (e4) at (5,4.5) {$\textcolor{blue}{\textcolor{blue}{\mathbf{e}_{4}}}$};  
	\node         (e12) at (-7.3,3) {$\mathbf{e}_{12}$};  
	\node         (e13) at (-5,3) {$\textcolor{blue}{\mathbf{e}_{13}}$};  
	\node         (e14) at (-3.5,3) {$\textcolor{blue}{\mathbf{e}_{14}}$};  
	\node         (e23) at (-2,3) {$\textcolor{blue}{\mathbf{e}_{23}}$};  
	\node         (e24) at (0,3) {$\textcolor{blue}{\mathbf{e}_{24}}$};  
	\node         (e34) at (2,3) {$\textcolor{blue}{\mathbf{e}_{34}}$};  
	\node         (e123) at (-8,1.5) {$\mathbf{e}_{123}$};  
	\node         (e124) at (-6.5,1.5) {$\textcolor{blue}{\mathbf{e}_{124}}$};  
	\node         (e134) at (-5,1.5) {$\textcolor{blue}{\mathbf{e}_{134}}$};  
	\node         (e234) at (-2,1.5) {$\textcolor{blue}{\mathbf{e}_{234}}$}; 
	\node         (e1234) at (-8,0) {$\textcolor{black}{\mathbf{e}_{1234}}$};  
	\draw[useful] (1.south) -- (e1.north east);
	\draw[useless] (1.south) -- (e2.north);
	\draw[useless] (1.south) -- (e3.north);
	\draw[useless] (1.south) -- (e4.north west);
	\draw[useful] (e1.south) -- (e12.north);
	\draw[useless] (e1.south) -- (e13.north);
	\draw[useless] (e1.south) -- (e14.north);
  \draw[useless] (e2.south) -- (e23.north);
  \draw[useless] (e2.south) -- (e24.north);
  \draw[useless] (e3.south) -- (e34.north);
  \draw[useful] (e12.south) -- (e123.north);
  \draw[useless] (e12.south) -- (e124.north);
  \draw[useless] (e13.south) -- (e134.north);
  \draw[useless] (e23.south) -- (e234.north);
  \draw[useful] (e123.south) -- (e1234.north);
  	\draw[black,rounded corners,-,color=gray!60] (-8.6,0.23) rectangle (-7.4,-0.23);
\end{tikzpicture}
\caption{~} 
\end{subfigure}
~\\
\begin{subfigure}{0.85\textwidth}
\begin{tikzpicture}[scale=0.60, every node/.style={scale=0.85}]
	\tikzset{primal/.style={->,>=latex,thin}}
	\tikzset{useful/.style={->,>=latex,thin,color=black}}
	\tikzset{useless/.style={dashed,->,>=latex,thin,color=blue}}
	\tikzset{uselessover/.style={dashed,->,>=latex,thin,color=forestGreen}}
	\node         (1) at (0,6) {$\mathbf{1}$};  
	\node         (e1) at (-5,4.5) {$\mathbf{e}_{1}$};  
	\node         (e2) at (-1,4.5) {$\mathbf{e}_{2}$};  
	\node         (e3) at (2,4.5) {$\textcolor{blue}{\mathbf{e}_{3}}$};  
	\node         (e4) at (5,4.5) {$\textcolor{blue}{\mathbf{e}_{4}}$};  
	\node         (e12) at (-7.3,3) {$\mathbf{e}_{12}$};  
	\node         (e13) at (-5,3) {$\mathbf{e}_{13}$};  
	\node         (e14) at (-3.5,3) {$\textcolor{blue}{\mathbf{e}_{14}}$};  
	\node         (e23) at (-2,3) {$\mathbf{e}_{23}$};  
	\node         (e24) at (0,3) {$\textcolor{blue}{\mathbf{e}_{24}}$};  
	\node         (e34) at (2,3) {$\textcolor{blue}{\mathbf{e}_{34}}$};  
	\node         (e123) at (-8,1.5) {$\mathbf{e}_{123}$};  
	\node         (e124) at (-6.5,1.5) {$\mathbf{e}_{124}$};  
	\node         (e134) at (-5,1.5) {$\mathbf{e}_{134}$};  
	\node         (e234) at (-2,1.5) {$\mathbf{e}_{234}$}; 
	\node         (e1234) at (-8,0) {$\textcolor{forestGreen}{\mathbf{e}_{1234}}$};  
	\draw[useful] (1.south) -- (e1.north east);
	\draw[useful] (1.south) -- (e2.north);
	\draw[useless] (1.south) -- (e3.north);
	\draw[useless] (1.south) -- (e4.north west);
	\draw[useful] (e1.south) -- (e12.north);
	\draw[useful] (e1.south) -- (e13.north);
	\draw[useless] (e1.south) -- (e14.north);
  \draw[useful] (e2.south) -- (e23.north);
  \draw[useless] (e2.south) -- (e24.north);
  \draw[useless] (e3.south) -- (e34.north);
  \draw[useful] (e12.south) -- (e123.north);
  \draw[useful] (e12.south) -- (e124.north);
  \draw[useful] (e13.south) -- (e134.north);
  \draw[useful] (e23.south) -- (e234.north);
  \draw[uselessover] (e123.south) -- (e1234.north);
  	\draw[black,rounded corners,-,color=gray!60] (-8.5,1.28) rectangle (-1.5,1.72);
\end{tikzpicture}
\caption{~} 
\end{subfigure}
~\\
\begin{subfigure}{0.85\textwidth}
\begin{tikzpicture}[scale=0.60, every node/.style={scale=0.85}]
	\tikzset{primal/.style={->,>=latex,thin}}
	\tikzset{useful/.style={->,>=latex,thin,color=black}}
	\tikzset{useless/.style={dashed,->,>=latex,thin,color=blue}}
	\tikzset{uselessover/.style={dashed,->,>=latex,thin,color=forestGreen}}
	\node         (1) at (0,6) {$\mathbf{1}$};  
	\node         (e1) at (-5,4.5) {$\mathbf{e}_{1}$};  
	\node         (e2) at (-1,4.5) {$\mathbf{e}_{2}$};  
	\node         (e3) at (2,4.5) {$\textcolor{black}{\mathbf{e}_{3}}$};  
	\node         (e4) at (5,4.5) {$\textcolor{blue}{\mathbf{e}_{4}}$};  
	\node         (e12) at (-7.3,3) {$\mathbf{e}_{12}$};  
	\node         (e13) at (-5,3) {$\mathbf{e}_{13}$};  
	\node         (e14) at (-3.5,3) {$\textcolor{black}{\mathbf{e}_{14}}$};  
	\node         (e23) at (-2,3) {$\mathbf{e}_{23}$};  
	\node         (e24) at (0,3) {$\textcolor{black}{\mathbf{e}_{24}}$};  
	\node         (e34) at (2,3) {$\textcolor{black}{\mathbf{e}_{34}}$};  
	\node         (e123) at (-8,1.5) {$\textcolor{forestGreen}{\mathbf{e}_{123}}$};  
	\node         (e124) at (-6.5,1.5) {$\textcolor{forestGreen}{\mathbf{e}_{124}}$};  
	\node         (e134) at (-5,1.5) {$\textcolor{forestGreen}{\mathbf{e}_{134}}$};  
	\node         (e234) at (-2,1.5) {$\textcolor{forestGreen}{\mathbf{e}_{234}}$}; 
	\node         (e1234) at (-8,0) {$\textcolor{forestGreen}{\mathbf{e}_{1234}}$};  
	\draw[useful] (1.south) -- (e1.north east);
	\draw[useful] (1.south) -- (e2.north);
	\draw[useful] (1.south) -- (e3.north);
	\draw[useless] (1.south) -- (e4.north west);
	\draw[useful] (e1.south) -- (e12.north);
	\draw[useful] (e1.south) -- (e13.north);
	\draw[useful] (e1.south) -- (e14.north);
  \draw[useful] (e2.south) -- (e23.north);
  \draw[useful] (e2.south) -- (e24.north);
  \draw[useful] (e3.south) -- (e34.north);
  \draw[uselessover] (e12.south) -- (e123.north);
  \draw[uselessover] (e12.south) -- (e124.north);
  \draw[uselessover] (e13.south) -- (e134.north);
  \draw[uselessover] (e23.south) -- (e234.north);
  \draw[uselessover] (e123.south) -- (e1234.north);
  	\draw[black,rounded corners,-,color=gray!60] (-7.7,2.8) rectangle (2.4,3.23);
\end{tikzpicture}
\caption{~} 
\end{subfigure}
~\\
\begin{subfigure}{0.85\textwidth}
\begin{tikzpicture}[scale=0.60, every node/.style={scale=0.85}]
	\tikzset{primal/.style={->,>=latex,thin}}
	\tikzset{useful/.style={->,>=latex,thin,color=black}}
	\tikzset{useless/.style={dashed,->,>=latex,thin,color=blue}}
	\tikzset{uselessover/.style={dashed,->,>=latex,thin,color=forestGreen}}
	\node         (1) at (0,6) {$\mathbf{1}$};  
	\node         (e1) at (-5,4.5) {$\mathbf{e}_{1}$};  
	\node         (e2) at (-1,4.5) {$\mathbf{e}_{2}$};  
	\node         (e3) at (2,4.5) {$\textcolor{black}{\mathbf{e}_{3}}$};  
	\node         (e4) at (5,4.5) {$\textcolor{black}{\mathbf{e}_{4}}$};  
	\node         (e12) at (-7.3,3) {$\textcolor{forestGreen}{\mathbf{e}_{12}}$};  
	\node         (e13) at (-5,3) {$\textcolor{forestGreen}{\mathbf{e}_{13}}$};  
	\node         (e14) at (-3.5,3) {$\textcolor{forestGreen}{\mathbf{e}_{14}}$};  
	\node         (e23) at (-2,3) {$\textcolor{forestGreen}{\mathbf{e}_{23}}$};  
	\node         (e24) at (0,3) {$\textcolor{forestGreen}{\mathbf{e}_{24}}$};  
	\node         (e34) at (2,3) {$\textcolor{forestGreen}{\mathbf{e}_{34}}$};  
	\node         (e123) at (-8,1.5) {$\textcolor{forestGreen}{\mathbf{e}_{123}}$};  
	\node         (e124) at (-6.5,1.5) {$\textcolor{forestGreen}{\mathbf{e}_{124}}$};  
	\node         (e134) at (-5,1.5) {$\textcolor{forestGreen}{\mathbf{e}_{134}}$};  
	\node         (e234) at (-2,1.5) {$\textcolor{forestGreen}{\mathbf{e}_{234}}$}; 
	\node         (e1234) at (-8,0) {$\textcolor{forestGreen}{\mathbf{e}_{1234}}$};  
	\draw[useful] (1.south) -- (e1.north east);
	\draw[useful] (1.south) -- (e2.north);
	\draw[useful] (1.south) -- (e3.north);
	\draw[useful] (1.south) -- (e4.north west);
	\draw[uselessover] (e1.south) -- (e12.north);
	\draw[uselessover] (e1.south) -- (e13.north);
	\draw[uselessover] (e1.south) -- (e14.north);
    \draw[uselessover] (e2.south) -- (e23.north);
    \draw[uselessover] (e2.south) -- (e24.north);  
    \draw[uselessover] (e3.south) -- (e34.north);
    \draw[uselessover] (e12.south) -- (e123.north);
    \draw[uselessover] (e12.south) -- (e124.north);
    \draw[uselessover] (e13.south) -- (e134.north);
    \draw[uselessover] (e23.south) -- (e234.north);
    \draw[uselessover] (e123.south) -- (e1234.north);
  	\draw[black,rounded corners,-,color=gray!60] (-5.3,4.32) rectangle (5.32,4.72);
\end{tikzpicture}
\caption{~} 
\end{subfigure}
\caption{\footnotesize Tree structure for some resulting multivectors of grade~$4$~(A), grade $3$~(B), grade $2$~(C), grade $1$~(D) in a $4$-dimensional vector space. Useless branches are depicted in \textcolor{forestGreen}{green} dashed arrows above the targeted multivector and in \textcolor{blue}{blue} below. The targeted nodes are surrounded by a black rectangle.}
\label{fig::necessaryTraversal}
\end{figure}
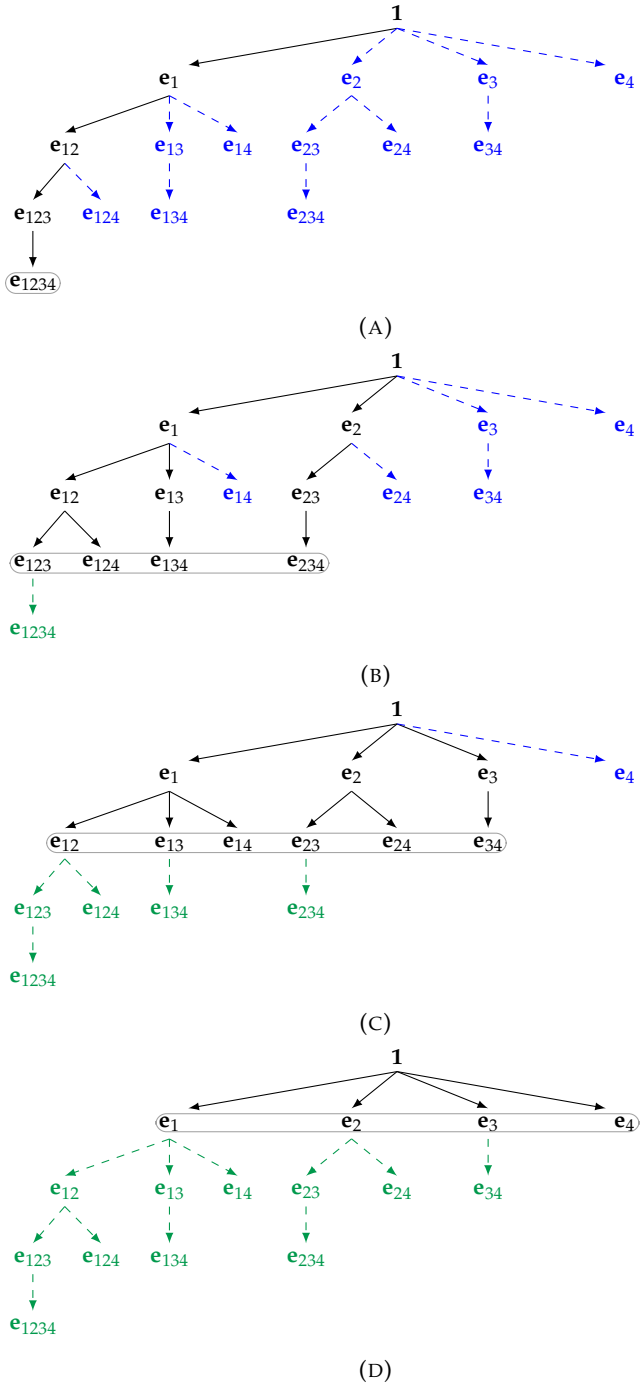

\begin{rem}[Recursive outer product in practice]
In practice, there are some obvious speed-up ways for Algorithm~\ref{algo:recursiveOuterProduct} on homogeneous multivectors, as stated in~\cite{breuils_garamon_2019}. The first way is to avoid recursive calls on nodes where the operand $\mathfrak{a}$ and $\mathfrak{b}$ lead to grade $g_a+g_b > g_c$. This is introduced in Definition~\eqref{eq:recursiveFormulaPrefixOuterproduct} as well as in Algorithm~\ref{algo:recursiveOuterProduct}, line~\ref{alg:recursiveOuterProduct_grade_optimization}.  
A more sophisticated speed-up way is to discard a recursive call on a branch that never reaches the grade of the considered multivector, as shown in blue dashed arrows in Figure~\ref{fig::necessaryTraversal}. These branch discard tests require only binary operators (very fast to compute) and can sometimes remove half of the recursive calls.
The pseudo-code of this speed-up way for the outer product is presented in Appendix~\ref{app:recursiveOuterProductPseudoCode}. The speed-up in running time is clear since it only removes some calls in the original algorithm, but the complexity study becomes more complicated.
\end{rem}
\newpage

\subsection{Recursive inner product}
\label{sec:recursive_inner_product}
As stated in Section~\ref{sec:innerProductProperties}, when $g_b > g_a$, the inner product is defined by the left contraction whereas it is by the right contraction when $g_b \leq g_a$. These two cases are thus treated separately.

As stated in the Section~\ref{sec:innerProductProperties}, the left contraction is a metric product and requires a metric to be defined. Let $\mathrm{M}_{d \times d}$ be the $d \times d$ symmetric matrix defining the vector inner product of the vector space of dimension $d$. In this context, we assume that the metric is diagonal. If not, the metric is assumed to be diagonalized. Thus, the metric will be only referred as its diagonal vector $\mathbf{m} = \mathrm{diag}(\mathrm{M}_{d \times d})$, where $\mathbf{m}(i) = \mathrm{M}_{d \times d}(i,i)$, such that
\begin{equation}
\begin{array}{r@{}l}
\mathbf{m}(1) &= \mathbf{e}_1 \cdot \mathbf{e}_1, \\
\mathbf{m}(2) &= \mathbf{e}_2 \cdot \mathbf{e}_2, \\
&~\vdots \\
\mathbf{m}(d) &= \mathbf{e}_d \cdot \mathbf{e}_d.
\end{array}
\end{equation} 
\begin{defn}[Recursive left contraction]
The construction of the recursive left contraction $\mathfrak{a} \rfloor \mathfrak{b}$ is defined as
\begin{equation}
\begin{array}{r@{}l}
 \text{at depth } n \\
\text{computation:~}  & \mathfrak{c}_{\lambda} \mathrel{+}= \mathfrak{a}_{\gamma} \rfloor \mathfrak{b}_{\delta}\text{,}\quad\text{if }|\delta| = g_b  \\
\text{recursive calls:~}  & \mathfrak{c}_{\lambda} =  \sum_{ i = \sigma}^{d} \mathbf{m}(i) \overline{\mathfrak{a}_{\gamma+i}} \rfloor \mathfrak{b}_{\delta+i} \text{, }\quad \hfill{\sigma \in [ \max( \lambda )+1, \cdots, d ] }\\
\text{recursive calls:~}  & \mathfrak{c}_{\lambda + \sigma} = \overline{\mathfrak{a}_{\gamma}} \rfloor \mathfrak{b}_{\delta+ \sigma} \text{, } \hfill{ \sigma \in [ \max( \lambda )+1, \cdots, d ] }
\end{array}
\label{eq:recursiveFormulaPrefixLeftContraction}
\end{equation} 
Note that the above recursive formula is equivalent to
\begin{equation}
\begin{array}{r@{}l}
 \text{at depth } n \\
\text{computation:~}  & \mathfrak{c}_{\lambda} \mathrel{+}= \mathfrak{a}_{\gamma} \rfloor \mathfrak{b}_{\delta}\text{,}\quad\text{if }|\delta| = g_b  \\
\text{recursive calls:~}  & \mathfrak{c}_{\lambda + \sigma} = \overline{\mathfrak{a}_{\gamma}} \rfloor \mathfrak{b}_{\delta+ \sigma } +  \mathbf{m}(\sigma) \overline{\mathfrak{a}_{\gamma+\sigma}} \rfloor \mathfrak{b}_{\delta+\sigma+\psi}  \text{, } \hfill{ \sigma \in [ \max( \lambda )+1, \cdots, d ] }, \\
& \hfill{ \psi \in [ \max( \lambda )+1, \cdots, d ]}
\end{array}
\label{eq:recursiveFormulaPrefixLeftContractionReviewed}
\end{equation} 

\end{defn}
We present in Algorithm~\ref{algo:recursiveLeftContraction} a simple and intuitive way to implement the recursive left contraction.  

\begin{algorithm}[!ht]
\DontPrintSemicolon
\Fn{ \funct{leftCont}}{
\KwIn{~$\mathfrak{a}_{\gamma}, \mathfrak{b}_{\delta}$: node of multivectors $\mathbf{A}$ and $\mathbf{B}$,\\
{\Indp \Indp $\mathfrak{c}_{\lambda}$: nodes of the resulting multivector $\mathbf{C}$\\}
{\Indp \Indp $\mathtt{complement}$: recursive value ( $\pm 1$).\\}
{\Indp \Indp $\mathtt{sign}$: recursive sign coefficient ( $\pm 1$).\\}
{\Indp \Indp $\mathbf{m}$: diagonal coefficients of the metric.\\}
}
\SetKw{and}{and}
\SetKw{In}{in}
\SetKw{Suchthat}{such that}
\If{$\vert\delta\vert = g_b$}{$\mathfrak{c}_{\lambda} \mathrel{+}= \mathtt{sign} \times \mathfrak{a}_{\gamma} \times \mathfrak{b}_{\delta}  $ \;} 
{
\ForEach{  $\sigma \in [ \max( \lambda )+1, \cdots, d ]$  } {   
 \textcolor{gray}{// $\mathbf{m}(i) \overline{\mathfrak{a}_{\gamma+ \sigma}} \rfloor \mathfrak{b}_{\delta+ \sigma}$}\; 
    \funct{leftCont}($\mathfrak{a}_{\gamma+ \sigma},\mathfrak{b}_{\delta + \sigma}, \mathfrak{c}_{\lambda},\mathbf{m}(\sigma) \times \mathtt{sign},-\mathtt{complement}$) \; 
    
}
\ForEach{  $\sigma \in [ \max( \lambda )+1, \cdots, d ]$  } {   
    \textcolor{gray}{// $\overline{\mathfrak{a}_{\gamma}} \rfloor \mathfrak{b}_{\delta+ \sigma}$}\;
    \funct{leftCont}($\mathfrak{a}_{\gamma},\mathfrak{b}_{\delta+\sigma},\mathfrak{c}_{\lambda+\sigma},\mathtt{sign},-\mathtt{complement} $)  \;
}
}
}
\textbf{First call:} \funct{leftCont}($\mathfrak{a}_{0},\mathfrak{b}_{0},\mathfrak{c}_{0},1,1)$
\caption{Pseudo-code of the recursive left contraction $\mathbf{C} = \mathbf{A} \rfloor \mathbf{B}$ }
\label{algo:recursiveLeftContraction}
\end{algorithm}

\begin{defn}[Recursive right contraction]
The construction of the recursive right contraction $\mathfrak{a} \lfloor \mathfrak{b}$ is defined as
\begin{equation}
\begin{array}{r@{}l}
 \text{at depth } n \\
\text{computation:~}   & \mathfrak{c}_{\lambda} \mathrel{+}= \mathfrak{a}_{\gamma} \lfloor \mathfrak{b}_{\delta}\text{,} \quad \text{if }|\delta| = g_a  \\
\text{recursive calls:~}  & \mathfrak{c}_{\lambda} =  \sum_{ i = \sigma}^{d} \mathbf{m}(i) \overline{\mathfrak{a}_{\gamma+i}} \lfloor \mathfrak{b}_{\delta+i} \text{, } \quad  \hfill{\sigma \in [ \max( \lambda )+1, \cdots, d ]} \\
\text{recursive calls:~}  & \mathfrak{c}_{\lambda + \sigma} = \mathfrak{a}_{\gamma+ \sigma} \lfloor \mathfrak{b}_{\delta}  \text{, } \hfill{ \sigma \in [ \max( \lambda )+1, \cdots, d ]} 
\end{array}
\label{eq:recursiveFormulaPrefixRightContraction}
\end{equation} 
\end{defn}
The algorithm of the recursive right contraction is presented in Algorithm~\ref{algo:recursiveRightContraction}.
\begin{algorithm}[!ht]
\DontPrintSemicolon
\Fn{ \funct{rightCont}}{
\KwIn{~$\mathfrak{a}_{\gamma}, \mathfrak{b}_{\delta}$: node of multivectors $\mathbf{A}$ and $\mathbf{B}$,\\
{\Indp \Indp $\mathfrak{c}_{\lambda}$: nodes of the resulting multivector $\mathbf{C}$\\}
{\Indp \Indp $\mathtt{complement}$: recursive value ( $\pm 1$).\\}
{\Indp \Indp $\mathtt{sign}$: recursive sign coefficient ( $\pm 1$).\\}
{\Indp \Indp $\mathbf{m}$: diagonal coefficients of the metric.\\}
}
\SetKw{and}{and}
\SetKw{In}{in}
\SetKw{Suchthat}{such that}
\If{$\vert\gamma\vert = g_a$}{$\mathfrak{c}_{\lambda} \mathrel{+}= \mathtt{sign} \times \mathfrak{a}_{\gamma} \times \mathfrak{b}_{\delta}  $ \;}
{
\ForEach{  $\sigma \in [ \max( \lambda )+1, \cdots, d ]$  } {   
 \textcolor{gray}{// $\mathbf{m}(i) \mathfrak{a}_{\gamma+ \sigma} \lfloor \mathfrak{b}_{\delta+ \sigma}$}\; 
    \funct{rightCont}($\mathfrak{a}_{\gamma+ \sigma},\mathfrak{b}_{\delta + \sigma}, \mathfrak{c}_{\lambda},\mathbf{m}(\sigma) \times \mathtt{sign},-\mathtt{complement}$) \; 
    
}
    \ForEach{  $\sigma \in [ \max( \lambda )+1, \cdots, d ]$  } {   
    \textcolor{gray}{// $\mathfrak{a}_{\gamma+\sigma} \lfloor \mathfrak{b}_{\delta}$}\;
    \funct{rightCont}($\mathfrak{a}_{\gamma+\sigma},\mathfrak{b}_{\delta},\mathfrak{c}_{\lambda+\sigma},\mathtt{complement}\times \mathtt{sign},\mathtt{complement} $)  \;
}
}
}
\textbf{First call:} \funct{rightCont}($\mathfrak{a}_{0},\mathfrak{b}_{0},\mathfrak{c}_{0},1,1$)
\caption{Pseudo-code of the recursive right contraction $\mathbf{C} = \mathbf{A} \lfloor \mathbf{B}$ }
\label{algo:recursiveRightContraction}
\end{algorithm}

\begin{thm}
\label{theorem:recursive_inner_product}
The complexity $c_\cdot^{\rm rec}$ of the recursive inner product $\mathbf{C} = \mathbf{A} \cdot \mathbf{B}$ between two homogeneous multivectors $\mathbf{A}$ and $\mathbf{B}$  of respective grade $g_a$ and~$g_b$, with resulting grade~$g_c = |g_a-g_b|$, is expressed as
\begin{equation}
c_\cdot^{\rm rec} =
\mathcal{O}\Bigg(\binom{d}{g_c} \binom{d-g_c}{\displaystyle \frac{g_a + g_b - g_c}{2}} \Bigg),
\label{eq:cplxtyRecInnerProdRecu}
\end{equation}
where $d$ is the dimension of the vector space.
\end{thm}
\begin{proof}
Lemma~\ref{lem:sameProductsForAllBrothers} still holds even for the recursive inner product. Namely, the siblings at any depth (grade) of the resulting prefix tree have the same number of inner products, i.e, the same number of recursive calls.
Furthermore, there are $\binom{d}{g_c}$ nodes of grade $g_c$ in the prefix tree represented in the $d$-dimensional vector space. The number of products of a given depth is thus the multiplication of the binomial coefficient and the number $n_{d-g_c,g_a}$ of inner  products per node of grade~$g_c$. The overall complexity is thus
\begin{equation}
    c_\cdot^{\rm rec} = \mathcal{O} \Bigg( \binom{d}{g_c} n_{d-g_c,g_a} \Bigg).
 \label{eq:outerprodComplexity_step1}
\end{equation}
To compute this number $n_{d-g_c,g_a}$ of recursive calls, let us focus on the evolution of grade $g_a$ with respect to a depth of recursion~$g_c$. The following proof is divided in two parts. The first part is dedicated to the case $g_a\geq g_b $ while the second part focuses on the case $g_a<g_b$.

In the first case, the considered product is the recursive left contraction, resulting in multivector $\mathfrak{c}=\mathfrak{a} \rfloor \mathfrak{b}$.
As stated in Equation~\eqref{eq:recursiveFormulaPrefixLeftContractionReviewed}, for a given grade~$g_c$ of the result, both recursive calls increase grade~$g_c$. The left-most term leaves grade~$g_a$ unchanged on one hand (and increases the grade of $\mathfrak{b}$):
\begin{equation}
n_{g_c,g_a} \rightarrow n_{g_c+1,g_a}
\label{eq:inner_recursive_part1}
\end{equation}
On the other hand, the second recursive call of Equation~\eqref{eq:recursiveFormulaPrefixLeftContractionReviewed} increases $g_a$ (and increases the grade of $\mathfrak{b}$).  
\begin{equation}
n_{g_c,g_a} \rightarrow n_{g_c+1,g_a+1}
\label{eq:inner_recursive_part2}
\end{equation}
By replacing $g_c$ by $d-g_c$ and summing Equations~\eqref{eq:inner_recursive_part1} and~\eqref{eq:inner_recursive_part2}, we obtain
\begin{equation}
    \begin{array}{ccc}
         n_{d-g_c-1,g_a+1} & = & n_{d-g_c,g_a} + n_{d-g_c,g_a+1}
    \end{array}.
    \label{eq:recursiveInnerProductCombined}
\end{equation}
In this context, the cases where either $d-g_c = g_a$ or $g_a = 0$ 
correspond to a final recursion condition, and thus we have $n_{d-g_c,g_a} = 1$. Therefore, Equation~\eqref{eq:recursiveInnerProductCombined} corresponds to the recursive definition of the binomial coefficient:
\begin{equation}
    n_{d-g_c,g_a} = 
    \binom{d-g_c-1}{g_a+1} = \binom{d-g_c}{g_a} + \binom{d-g_c}{g_a+1}.
\end{equation}
Hence, the complexity $c_\rfloor^{\rm rec}$ of the recursive left contraction is
\begin{equation}
    c_\rfloor^{\rm rec} = 
    \mathcal{O}\Bigg(\binom{d}{g_c} \binom{d-g_c}{g_a}  \Bigg).
\end{equation}
In a similar manner, we have the complexity $c_\lfloor^{\rm rec}$ of the recursive right contraction is
\begin{equation}
    c_\lfloor^{\rm rec} = 
    \mathcal{O}\Bigg(\binom{d}{g_c} \binom{d-g_c}{g_b}  \Bigg).
\end{equation}
Accordingly, the complexity of the recursive inner product is
\begin{equation}
c_\cdot^{\rm rec} =
c_\rfloor^{\rm rec} + c_\lfloor^{\rm rec} = \mathcal{O}\Bigg(\binom{d}{g_c} \binom{d-g_c}{\displaystyle \frac{g_a + g_b - g_c}{2}}\Bigg) .
\end{equation}
\end{proof}

\subsection{Recursive geometric product}
Similarly to the two other products, let us start with the definition of the recursive geometric product. 

\begin{defn}
Given two homogeneous multivectors $\mathbf{A}$ and $\mathbf{B}$ and the set $\mathcal{I} = \{|g_a - g_b |,|g_a - g_b |+2, \dots , g_a + g_b \}$, where $g_a$ and~$g$ are respectively the grade of~$\mathbf{A}$ and~$\mathbf{B}$, the recursive geometric product is expressed as
\begin{equation}
\begin{array}{r@{}l}
 \text{at depth } n \\
\text{computation:~}   & \mathfrak{c}_{\lambda} \mathrel{+}= \mathfrak{a}_{\gamma} \times \mathfrak{b}_{\delta}\text{,}\quad\text{if }|\lambda| \in \mathcal{I},|\gamma| = g_a   \\
\text{recursive calls:~}  & \mathfrak{c}_{\lambda} =  \sum_{ i = \sigma}^{d} \mathbf{m}(i) \overline{\mathfrak{a}_{\gamma+i}} * \mathfrak{b}_{\delta+i} \hfill{\sigma \in [ \max( \lambda )+1, \cdots, d ]} \\
\text{recursive calls:~}  & \mathfrak{c}_{\lambda + \sigma} = \mathfrak{a}_{\gamma+ \sigma} * \mathfrak{b}_{\delta} + \overline{\mathfrak{a}_{\gamma}} * \mathfrak{b}_{\delta+ \sigma}   \text{, } \quad\hfill{\sigma \in [ \max( \lambda )+1, \cdots, d ]}
\end{array}.
\label{eq:recursiveFormulaPrefixGeoproduct}
\end{equation}
\end{defn}
The pseudo-code for this definition is presented in Algorithm~\ref{algo:recursiveGeometricProduct}.
\begin{algorithm}[!ht]
\DontPrintSemicolon
\Fn{ \funct{geoProduct}}{
\KwIn{~$\mathfrak{a}_{\gamma}, \mathfrak{b}_{\delta}$: node of multivectors $\mathbf{A}$ and $\mathbf{B}$,\\
{\Indp \Indp $\mathfrak{c}_{\lambda}$: nodes of the resulting multivector $\mathbf{C}$\\}
{\Indp \Indp $\mathtt{complement}$: recursive value ( $\pm 1$).\\}
{\Indp \Indp $\mathtt{sign}$: recursive sign coefficient ( $\pm 1$).\\}
{\Indp \Indp $\mathbf{m}$: coefficients of the metric.\\}
{\Indp \Indp $\mathcal{I} = \{|g_a - g_b |,|g_a - g_b |+2, \dots , g_a + g_b  \} $.\\}
}
\SetKw{and}{and}
\SetKw{In}{in}
\SetKw{Suchthat}{such that}
\If{$\vert \lambda \vert \in \mathcal{I}$ \and $\vert \gamma \vert = g_a$ }{$\mathfrak{c}_{\lambda} \mathrel{+}= \mathtt{sign} \times \mathfrak{a}_{\gamma} \times \mathfrak{b}_{\delta}  $ \;}

{
\ForEach{  $\sigma \in [ \max( \lambda )+1, \cdots, d ]$  } {   
 \textcolor{gray}{// $\mathbf{m}(i) \mathfrak{a}_{\gamma+ \sigma} * \mathfrak{b}_{\delta+ \sigma}$}\; 
    \funct{geoProduct}($\mathfrak{a}_{\gamma+ \sigma},\mathfrak{b}_{\delta + \sigma}, \mathfrak{c}_{\lambda},\mathbf{m}(\sigma) \times \mathtt{sign},-\mathtt{complement}$) \; 
}
\ForEach{  $\sigma \in [ \max( \lambda )+1, \cdots, d ]$  } {   
     \textcolor{gray}{// $\mathfrak{a}_{\gamma+\sigma} *  \mathfrak{b}_{\delta}$}\;
    \funct{geoProduct}($\mathfrak{a}_{\gamma+\sigma},\mathfrak{b}_{\delta},\mathfrak{c}_{\lambda+\sigma},\mathtt{complement}\times \mathtt{sign},\mathtt{complement} $)  \;
    \textcolor{gray}{// $\overline{\mathfrak{a}_{\gamma}} * \mathfrak{b}_{\delta+\sigma}$}\;
    \funct{geoProduct}($\mathfrak{a}_{\gamma},\mathfrak{b}_{\delta+\sigma},\mathfrak{c}_{\lambda+\sigma},\mathtt{sign},-\mathtt{complement} $)  \;    
}
}
}
\textbf{First call:} \funct{geoProduct}($\mathfrak{a}_{0},\mathfrak{b}_{0},\mathfrak{c}_{0},1,1)$
\caption{Pseudo-code of the recursive geometric product $\mathbf{C} = \mathbf{A} * \mathbf{B}$}
\label{algo:recursiveGeometricProduct}
\end{algorithm}

\begin{thm}
The complexity $c_{*}^{\rm rec}$ of the recursive geometric product $\mathbf{C} = \mathbf{A} * \mathbf{B}$ between two homogeneous multivectors $\mathbf{A}$ and $\mathbf{B}$  of respective grade $g_a$ and~$g_b$, with resulting grade~$g_c \in \mathcal{I} = \{|g_a - g_b |,|g_a - g_b |+2, \dots , g_a + g_b  \} $, is expressed as
\begin{equation}
c_*^{\rm rec} = \mathcal{O}\Bigg(\binom{d}{g_c}  \binom{g_c}{\displaystyle\frac{g_a - g_b + g_c}{2}} \binom{d-g_c}{\displaystyle\frac{g_a + g_b - g_c}{2}} \Bigg),
\label{eq:cplxtyRecGeoProdRecu}
\end{equation}
where $d$ is the dimension of the vector space.
\end{thm}

\begin{proof}
This proof is split into three parts, each of which is dedicated to one term in Equation~\eqref{eq:cplxtyRecGeoProdRecu}. As for the outer and inner products, the number of recursive calls remains the same for any nodes of grade $g_c$.  Moreover, there are $\binom{d}{g_c}$ products for each node of grade $g_c$ of the resulting multivector in the $d$-dimensional vector space. Let us denote by $n_{g_a,g_b,g_c}$ the number of recursive calls with respect to grades $g_a,g_b$, and $g_c$. The overall complexity is then
\begin{equation}
    c_*^{\rm rec} = \mathcal{O} \Bigg( \binom{d}{g_c} n_{g_a,g_b,g_c} \Bigg).
\end{equation}

Let us now reason the recursive formula of Equation~\eqref{eq:recursiveFormulaPrefixGeoproduct}. We remark that the recursive calls that increase the grade of the resulting multivector are those coming only from the outer product of Equation~\eqref{eq:recursiveFormulaPrefixOuterproduct}, corresponding to the last recursive call of Equation~\eqref{eq:recursiveFormulaPrefixGeoproduct}.
As previously studied in Equation~\eqref{eq:cplxtyRecOuterProd}, for each possible grade of $\mathfrak{c}$, the number of calls associated with the recursive outer product is 
\begin{equation}
    \binom{g_c}{g_a} = \binom{g_c}{\displaystyle\frac{g_a - g_b + (g_a + g_b)}{2}} = \binom{d}{\displaystyle\frac{g_a - g_b + g_c}{2}} .
\end{equation}
Then, for any of the recursive outer product calls of the recursive geometric product, the recursive calls can be split into  
\begin{equation}
\begin{array}{r@{}l}
 \text{at depth } n \\
\text{computation:~}   & \mathfrak{c}_{\lambda} \mathrel{+}= \mathfrak{a}_{\gamma} \times \mathfrak{b}_{\delta}\text{,}\quad\text{if }|\lambda| \in \mathcal{I}, \vert \gamma \vert = g_a  \\
\text{recursive calls:~}  & \mathfrak{c}_{\lambda} =  \sum_{ i = \sigma}^{d} \mathbf{m}(i) \overline{\mathfrak{a}_{\gamma+i}} * \mathfrak{b}_{\delta+i} ~~~ \sigma \in [ \max( \lambda )+1, \cdots, d ] \\
\text{recursive calls:~}  & \mathfrak{c}_{\lambda + \sigma} = \mathfrak{a}_{\gamma+ \sigma} * \mathfrak{b}_{\delta}  \text{, } ~~~ \sigma \in [ \max( \lambda )+1, \cdots, d ] \end{array}
\label{eq:cloneOfRecursiveRightContraction}
\end{equation} 
and 
\begin{equation}
\begin{array}{r@{}l}
 \text{at depth } n \\
\text{computation:~}   & \mathfrak{c}_{\lambda} \mathrel{+}= \mathfrak{a}_{\gamma} \times \mathfrak{b}_{\delta}\text{,}\quad\text{if }|\lambda| \in \mathcal{I}, \vert \gamma \vert = g_a\\
\text{recursive calls:~}  & \mathfrak{c}_{\lambda} =  \sum_{ i = \sigma}^{d} \mathbf{m}(i) \overline{\mathfrak{a}_{\gamma+i}} * \mathfrak{b}_{\delta+i} ~~~ \sigma \in [ \max( \lambda )+1, \cdots, d ] \\
\text{recursive calls:~}  & \mathfrak{c}_{\lambda + \sigma} = \overline{\mathfrak{a}_{\gamma}} * \mathfrak{b}_{\delta+ \sigma}   \text{, } ~~~ \sigma \in [ \max( \lambda )+1, \cdots, d ] 
\end{array}.
\label{eq:cloneOfRecursiveLeftContraction}
\end{equation} 

We recognize the recursive right contraction of Equation~\eqref{eq:recursiveFormulaPrefixRightContraction} in Equation~\eqref{eq:cloneOfRecursiveRightContraction} whereas Equation~\eqref{eq:cloneOfRecursiveLeftContraction} corresponds to the recursive left contraction of Equation~\eqref{eq:recursiveFormulaPrefixLeftContraction}. 
This indicates that for each recursive outer product call, recursive inner product calls are executed. Following the arguments of Theorem~\ref{theorem:recursive_inner_product}, we see that the number of required recursive calls is
\begin{equation}
    n_{g_a,g_b,g_c} = \binom{g_c}{\displaystyle\frac{g_a - g_b + g_c}{2}} \binom{d-g_c}{\displaystyle\frac{g_a + g_b - g_c}{2}}
\end{equation}
for any grade $g_c \in \mathcal{I}$.
By merging the above arguments, we have Equation~\eqref{eq:cplxtyRecGeoProdRecu}.
\end{proof}

\section{Discussion}\label{sec:discussion}
The motivation of this paper is to study the complexity of geometric algebra products over homogeneous multivectors. Sections~\ref{sec:outer}, \ref{sec:inner}, and \ref{sec:geometric} focus on the outer product, inner product, and geometric product respectively. Table~\ref{tab:summarizeArithmeticOpe} summarizes the required number of operation for each product according to the dimension of the considered vector space and the grades of the homogeneous multivectors used for the products.

This product complexity study naturally raises a subsidiary study about effective implementations of geometric algebra products. A first approach consists of pre-computing the products for a given algebra. The resulting code always reaches the best complexity for homogeneous multivectors. A second approach consists of a syntax simplification of geometric algebra expression. In principle, this technique also reaches the best complexity and can sometimes perform even better.

For both the approaches, one may also wonder if the pre-computation process is optimal. This question especially makes sense for meta-programming when the compilation time is important or for the cases where the products are computed on the fly for higher dimensional vector spaces. 

In the case where the pre-computation is performed from existing product tables, each product between $\mathbf{A}$ and $\mathbf{B}$ requires to read all the entries of the table for the grades ($g_a$,$g_b)$. Some entries will lead to a pre-computed product when some other will just result in zero. The complexity is then in $\mathcal{O}(\binom{d}{g_a}\binom{d}{g_b})$ for every product. Thus, the table approach is optimal for the geometric product but neither for the outer product nor for the inner product.

Another solution is to pre-compute the product using the XOR operators~\cite{Eid2018}. Then for the $\binom{d}{g_a}\binom{d}{g_b}$ possible products, required are to check whether the result is non-zero and to compute the resulting sign. Note that the sign computation is in $\mathcal{O}(d)$, see Section~\ref{section:introduction} or~\cite{Eid2018} for further details. The resulting complexity becomes $\mathcal{O}(d \times \binom{d}{g_a}\binom{d}{g_b})$.

Finally, the product pre-computation (or computation on the fly) can be performed by the recursive form presented in Section~\ref{sec:recursiveApproach}. For each product, this method reaches the optimal complexity, as stated in Table~\ref{tab:summarizeComplexity}.

\section{Conclusion}
We gave a study on the number of arithmetic operations required for the outer, inner, and geometric products of geometric algebra for any full homogeneous multivectors with only non-zero elements. 
This study allowed to prove that there exists an approach that reaches the equivalent complexity for each product. 
As a perspective of this paper, we would focus on the computational complexity of products with more than two homogeneous multivectors.

\bibliographystyle{acm}
\bibliography{main}

\appendix

\section{Proof of Lemma~\ref{lem:sameProductsForAllBrothers}}
\label{proof:sameOuterProductsEachNode}
Let us prove it by induction using the recursive formula~\eqref{eq:recursiveFormulaPrefixOuterproduct}. The base case is $g_c = 0$. The recursive formula~\eqref{eq:recursiveFormulaPrefixOuterproduct} yields:
\begin{equation}
\begin{array}{r@{}l}
\text{at depth } 0 \quad\\
\text{computation:~}   & ~~\mathfrak{c}_{\lambda}  \mathrel{+}= \mathfrak{a}_{\gamma} \wedge \mathfrak{b}_{\delta},\quad\text{if }|\lambda| = g_c  \\ 
\text{recursive calls:~}  & \mathfrak{a}_{\sigma} \wedge \mathfrak{b}_{0} + \overline{\mathfrak{a}_{0}} \wedge \mathfrak{b}_{\sigma} \text{, } ~~~ \sigma \in [ \max( \lambda )+1, \cdots, d ] 
\end{array}.
\end{equation} 
We remark that each node of the resulting outer product prefix tree of grade $1$ is $2$. Then, all the siblings of grade $1$ induce the same number of products.  

Let us assume that the proposition holds for a given grade of $\mathfrak{c}$, called $k_c \in \mathbb{N}$. Then the recursive products associated with any nodes $\mathfrak{c}_{\lambda}$ of grade $k_c$ can be 
seen as the sum of products with the same number of terms. For any node, each single product can be written as
\begin{equation}
\mathfrak{c}_{\lambda} = \mathfrak{a}_{\mu} \wedge \mathfrak{b}_{\nu}.  
\end{equation}
This product expand at the grade of $k_c + 1$ is as follows.
\begin{equation}
\begin{array}{r@{}l}
\text{at depth } k_c +1 \quad\\
\text{computation:~}   & ~~\mathfrak{c}_{\lambda}  \mathrel{+}= \mathfrak{a}_{\gamma} \wedge 
\mathfrak{b}_{\delta},\quad\text{if }|\lambda| = g_c  \\
\text{recursive calls:~} & \mathfrak{c}_{\lambda + \sigma} = \mathfrak{a}_{\mu+ \sigma} \wedge \mathfrak{b}_{\nu} + \overline{\mathfrak{a}_{\mu}} \wedge \mathfrak{b}_{\nu+ \sigma} \text{, } ~~~ \sigma \in [ \max( \lambda )+1, \cdots, d ] 
\end{array}
\end{equation} 
Again, we remark that for any nodes of $\mathfrak{c}$ of grade $k_c + 1$, the number of products remains the same. Thus, by induction, the number of outer products remains the same for any node of the resulting prefix tree having the same grade (depth).

\section{Pseudo-codes of the recursive products}
In the optimized pseudo-code, the indices of the basis blades are represented with a binary label. This binary label is useful to optimize paths in the prefix tree.
The binary label of a node is recursively computed using the binary label of its parent node. A node with binary label $\mathtt{u}$ has its first child binary label computed by
\begin{equation} \label{eq:child}
\mathrm{child\_label}(\mathtt{u},\mathtt{msb}) = \mathtt{u} + \mathtt{msb},
\end{equation}
where $+$ is the binary addition and $\mathtt{msb}$ is the binary label of the basis vector "added" to the basis blade by the outer product. So, $\mathtt{msb}$ contains only a single bit set to $1$. Note that this bit set to $1$ in $\mathtt{msb}$ cannot be a bit already set to $1$ in $\mathtt{u}$, otherwise the parent node and its child would have the same grade.

The contribution of $\mathtt{msb}$ is the most significant bit of $\mathrm{child\_label}(\mathtt{label},\allowbreak \mathtt{msb})$, i.e., the first bit to 1 encountered while reading the binary label from the left, which corresponds to the position of the $1$-bit of $\mathtt{msb}$. 

We show the pseudo-code of the optimized outer product with the definition of these functions in Algorithm~\ref{algo:recursiveOuterProductOptimized}.
In this algorithm, \funct{labelToMsb}$(\mathtt{label})$ computes $\mathtt{msb}$, the most significant bit from the considered label, i.e. the first $\mathtt{1}$ encountered in the binary word $\mathtt{label}$ when reading from left to right. 

\label{app:recursiveOuterProductPseudoCode}
\begin{algorithm}[!ht]
\DontPrintSemicolon
\Fn{ \funct{gradeKReachable}}{
\KwIn{$\mathtt{label}$: the recursive position \\
{\Indp \Indp  $\mathtt{msb}$: a label of the last traversed vector\\} 
{\Indp \Indp  $k$: the considered grade.\\}
}
\SetKw{and}{and}
\SetKw{In}{in}
\SetKw{Suchthat}{such that}
$\mathit{labelChildK} \leftarrow \mathtt{label} + \mathtt{msb} (2^{k-grade(\mathtt{label})}-1)$ \;
\Return $\mathtt{labelChildK}  < 2^{d}$\;
}

\Fn{ \funct{outer}}{
\KwIn{~$A, B$: two multivectors,\\
{\Indp \Indp $C$: resulting multivector,\\}
{\Indp \Indp $k_a$, $k_b$ and $k_c$: the respective grade of each multivector.\\}
{\Indp \Indp $\mathtt{label}_{a},\mathtt{label}_{b}, \mathtt{label}_{c}$: recursive position on each tree. \\} 
{\Indp \Indp $\mathtt{sign}$: recursive sign index.\\}
{\Indp \Indp $\mathtt{complement}$: recursive value ( $\pm 1$).\\} 
}
\SetKw{and}{and}
\SetKw{In}{in}
\SetKw{Suchthat}{such that}
\eIf(\hfill \textcolor{gray}{// end of recursion}){\funct{grade}$(\mathtt{label}_{c}) == k_c$}{
$C[\mathtt{label}_{c}] += \mathtt{sign} \times A[\mathtt{label}_{a}] \times B[\mathtt{label}_{b}]$ \;
}(\hfill \textcolor{gray}{// recursive calls}){
$\mathtt{msb}_{a} =$ \funct{labelToMsb}$(\mathtt{label}_{a})$ \;
$\mathtt{msb}_{b} =$ \funct{labelToMsb}$(\mathtt{label}_{b})$ \;
$\mathtt{msb}_{c} =$ \funct{labelToMsb}$(\mathtt{label}_{c})$ \;
\ForEach{$\mathtt{msb}$ \Suchthat \funct{gradeKReachable}$(k_c,\mathtt{msb},\mathtt{label}_c) \mathtt{==true}$} {   
\color{black}
$\mathtt{label} = \mathtt{label}_c + \mathtt{msb}$ \;
\If{\funct{gradeKReachable}$(k_a,\mathtt{msb},\mathtt{label}_a)$}{
{\funct{outer}($A,B,C,k_a,k_b,k_c, \mathtt{label}_{a}+\mathtt{msb},\mathtt{label}_{b},\mathtt{label}, \mathtt{sign} \times \mathtt{complement},\mathtt{complement}$)} \;
}
\If{\funct{gradeKReachable}$(k_b,\mathtt{msb},\mathtt{label}_b)$}{
{\funct{outer}($A,B,C,k_a,k_b,k_c, \mathtt{label}_{a}, \mathtt{label}_{b} + \mathtt{msb}, \mathtt{label}, \mathtt{sign}, -\mathtt{complement} $)} \;
}
}
}
}
\caption{Recursive outer product $C = A \wedge B$ }
\label{algo:recursiveOuterProductOptimized}
\end{algorithm}

We also give the pseudo-codes of the optimized left contraction, right contraction, and geometric product
in Algorithms~\ref{algo:recursiveLeftContractionProductOptimized}, \ref{algo:recursiveRightContractionProductOptimized}, and~\ref{algo:recursiveGeometricProductOptimized}, respectively. 
The functions called inside these pseudo-codes are the same as those in Algorithm~\ref{algo:recursiveOuterProductOptimized}. 

\begin{algorithm}[!ht]
\DontPrintSemicolon
\Fn{ \funct{leftcont}}{
\KwIn{~$A, B$: two multivectors.\\
{\Indp \Indp $C$: resulting multivector.\\}
{\Indp \Indp $k_a$, $k_b$ and $k_c$: respective grade of each multivector.\\}
{\Indp \Indp $\mathtt{label}_{a},\mathtt{label}_{b}, \mathtt{label}_{c}$: recursive position on each tree. \\} 
{\Indp \Indp $\mathtt{sign}$: a recursive sign index.\\}
}
{\Indp \Indp $\mathtt{complement}$: recursive value ($\pm 1$).\\}
{\Indp \Indp $\mathbf{m}$: vectors representing the metric diagonal matrix.\\}
\SetKw{and}{and}
\SetKw{In}{in}
\SetKw{Suchthat}{such that}
\eIf(\hfill \textcolor{gray}{// end of recursion}){\funct{grade}$(\mathtt{label}_{b}) == k_b$}{
$C[\mathtt{label}_{c}] += \mathbf{m} \times \mathtt{sign} \times A[\mathtt{label}_{a}] \times B[\mathtt{label}_{b}]$ \;
}(\hfill \textcolor{gray}{// recursive calls}){
$\mathtt{msb}_{a} =$ \funct{labelToMsb}$(\mathtt{label}_{a})$ \;
$\mathtt{msb}_{b} =$ \funct{labelToMsb}$(\mathtt{label}_{b})$ \;
$\mathtt{msb}_{c} =$ \funct{labelToMsb}$(\mathtt{label}_{c})$ \;
\ForEach{$\mathtt{msb}$ \Suchthat \funct{gradeKReachable}$(k_b,\mathtt{msb},\mathtt{label}_b) \mathtt{==true}$} {
$\mathtt{label} = \mathtt{label}_b + \mathtt{msb}$ \;
\If{\funct{gradeKReachable}$(k_a$, $\mathtt{msb}$, $\mathtt{label}_a)$}{
\funct{leftcont}($A$, $B$, $C$, $k_a$, $k_b$, $k_c$, $ \mathtt{label}_{a}+\mathtt{msb}$, $\mathtt{label}$, $\mathtt{label}_{c}$, $  \mathtt{sign} \times \mathtt{complement}$, $-\mathtt{complement}$, 
$\mathtt{metric} \times \mathbf{m}(grade(\mathtt{label}_b))$) \;
}
\If{\funct{gradeKReachable}$(k_c,\mathtt{msb},\mathtt{label}_c)$}{
\funct{leftcont}($A,B,C,k_a,k_b,k_c, \mathtt{label}_{a}, \mathtt{label}, \mathtt{label}_c + msb,  \mathtt{sign},-\mathtt{complement},\mathtt{metric}) $) \;
}
}
}
}
\caption{Recursive left contraction product $C = A  \rfloor B$}
\label{algo:recursiveLeftContractionProductOptimized}
\end{algorithm}


\begin{algorithm}[!ht]
\DontPrintSemicolon
\Fn{ \funct{rightcont}}{
\KwIn{~$\mathbf{A}, \mathbf{B}$: two multivectors.\\
{\Indp \Indp $\mathbf{C}$: resulting multivector.\\}
{\Indp \Indp $k_a$, $k_b$ and $k_c$: respective grade of each multivector.\\}
{\Indp \Indp $\mathtt{label}_{a},\mathtt{label}_{b}, \mathtt{label}_{c}$: recursive position on each tree. \\} 
{\Indp \Indp $\mathtt{sign}$: recursive sign index.\\}
}
{\Indp \Indp $\mathtt{complement}$: recursive value ($\pm 1$).\\}
{\Indp \Indp $\mathtt{metric}$: coefficients related to the metric.\\}
\SetKw{and}{and}
\SetKw{In}{in}
\SetKw{Suchthat}{such that}
\eIf(\hfill \textcolor{gray}{// end of recursion}){\funct{grade}$(\mathtt{label}_{b}) == k_b$}{
$\mathbf{C}[\mathtt{label}_{c} += \mathtt{metric} \times \mathtt{sign} \times \mathbf{A}[\mathtt{label}_{a}] \times \mathbf{B}[\mathtt{label}_{b}]$ \;
}(\hfill \textcolor{gray}{// recursive calls}){
$\mathtt{msb}_{a} =$ \funct{labelToMsb}$(\mathtt{label}_{a})$ \;
$\mathtt{msb}_{b} =$ \funct{labelToMsb}$(\mathtt{label}_{b})$ \;
$\mathtt{msb}_{c} =$ \funct{labelToMsb}$(\mathtt{label}_{c})$ \;
\ForEach{$\mathtt{msb}$ \Suchthat \funct{gradeKReachable}$(k_a,\mathtt{msb},\mathtt{label}_a) \mathtt{==true}$} {
$\mathtt{label} = \mathtt{label}_a + \mathtt{msb}$ \;
\If{\funct{gradeKReachable}$(k_b,\mathtt{msb},\mathtt{label}_b)$}{
\funct{rightcont}($\mathbf{A}$, $\mathbf{B}$, $\mathbf{C}$, $k_a$, $k_b$,$k_c$,  $\mathtt{label}$, $\mathtt{label}_{b}+\mathtt{msb}$, $\mathtt{label}_{c}$,   $\mathtt{sign} \times \mathtt{complement}$, $-\mathtt{complement}$, $\mathtt{metric} \times \mathbf{m}(grade(\mathtt{label}_b))$) \;
}
\If{\funct{gradeKReachable}$(k_c,\mathtt{msb},\mathtt{label}_c)$}{
\funct{rightcont}($\mathbf{A},\mathbf{B},\mathbf{C},k_a,k_b,k_c, \mathtt{label}, \mathtt{label}_{b}, \mathtt{label}_c + msb,  \mathtt{sign},-\mathtt{complement},\mathtt{metric}) $) \;
}
}
}
}
\caption{Recursive right contraction product $C = A  \lfloor B$}
\label{algo:recursiveRightContractionProductOptimized}
\end{algorithm}


\begin{algorithm}[!htp]
\DontPrintSemicolon
\Fn{ \funct{geometric}}{
\KwIn{~$\mathbf{A}, \mathbf{B}$: two multivectors.\\
{\Indp \Indp $\mathbf{C}$: resulting multivector.\\}
{\Indp \Indp $k_a$, $k_b$ and $k_c$: respective grade of each multivector.\\}
{\Indp \Indp $\mathtt{label}_{a},\mathtt{label}_{b}, \mathtt{label}_{c}$:  recursive position on each tree. \\} 
{\Indp \Indp $\mathtt{sign}$: a recursive sign index.\\}
{\Indp \Indp $\mathtt{complement}$: recursive value ($\pm 1$).\\}
{\Indp \Indp $\mathtt{metric}$:  coefficients related to the metric.\\}
{\Indp \Indp $\mathtt{depth}$: current depth in the prefix tree.\\}
}
\SetKw{and}{and}
\SetKw{In}{in}
\SetKw{Suchthat}{such that}
\eIf{
\funct{grade}$(\mathtt{label}_{b}) == k_b$ \and \funct{grade}$(\mathtt{label}_{a}) == k_a$}{
$\mathbf{C}[\mathtt{label}_{c}] += \mathtt{metric} \times \mathtt{sign} \times \mathbf{A}[\mathtt{label}_{a}] \times \mathbf{B}[\mathtt{label}_{b}]$ \;
\tcp*[f]{\color{gray} end of recursion}
}{
$\mathtt{msb}_{a} =$ \funct{labelToMsb}$(\mathtt{label}_{a})$ \;
$\mathtt{msb}_{b} =$ \funct{labelToMsb}$(\mathtt{label}_{b})$ \;
$\mathtt{msb}_{c} =$ \funct{labelToMsb}$(\mathtt{label}_{c})$ \;

\For{ $\mathtt{i}  $ \In $ 2^{\mathtt{depth}}, 2^{\mathtt{depth}+1},\cdots, 2^{d-1} $ } {
\color{black}
\If{\funct{gradeKReachable}$(k_b,\mathtt{i},\mathtt{label}_b)$}{
\If{\funct{gradeKReachable}$(k_a,\mathtt{i},\mathtt{label}_a)$}{
\funct{geometric}\big($\mathbf{A},\mathbf{B},\mathbf{C},k_a,k_b,k_c, \mathtt{label}_{a}+\mathtt{i},\mathtt{label}_{b}+\mathtt{i},\mathtt{label}_{c},  \mathtt{sign} \times \mathtt{complement}, -\mathtt{complement},\mathtt{metric} \times \mathbf{m}(\mathtt{i}),\mathtt{depth}+1)$\big) \;
}
}
\If{\funct{gradeKReachable}$(k_a,\mathtt{i},\mathtt{label}_a)$}{
\funct{geometric}\big($\mathbf{A},\mathbf{B},\mathbf{C},k_a,k_b,k_c, \mathtt{label}, \mathtt{label}_{b}, \mathtt{label}_c + msb, \mathtt{sign}\times\mathtt{complement}, \mathtt{complement}, \mathtt{metric}, \mathtt{depth}+1) $\big) \;
}

\If{\funct{gradeKReachable}$(k_b,\mathtt{i},\mathtt{label}_b)$}{
\funct{geometric}\big($\mathbf{A},\mathbf{B},\mathbf{C},k_a,k_b,k_c, \mathtt{label}_{a}, \mathtt{label}_{b}+\mathtt{i}, \mathtt{label}_{c}+\mathtt{i},  \mathtt{sign},-\mathtt{complement},\mathtt{metric}),\mathtt{depth}+1 $\big) \;
}
}
}
}
\caption{Recursive geometric product $\mathbf{C} = \mathbf{A}* \mathbf{B}$ }
\label{algo:recursiveGeometricProductOptimized}
\end{algorithm}

\end{document}